\documentclass[journal]{IEEEtran}
\usepackage[font=footnotesize,labelfont=bf]{caption}
\usepackage{amsmath}
\usepackage{booktabs}
\usepackage[inline]{enumitem}
\usepackage[export]{adjustbox}
\usepackage{wrapfig}
\usepackage{subfigure}
\usepackage{amssymb,amsfonts} % Add by Xiaofan
\usepackage{amsthm} % Add by Xiaofan
\usepackage{commath} % Add by Xiaofan
\usepackage{algorithmic}
\usepackage{mathtools, nccmath}% Add by Xiaofan
\usepackage{array}
\usepackage{textcomp}
\usepackage{xcolor}
\usepackage{comment}  % Added by Al
\usepackage{bm} % Added by ATA
\usepackage[final]{pdfpages}

\newlist{steps}{enumerate}{1}
\setlist[steps, 1]{label = Step \arabic*:}

\usepackage{setspace}

\newcommand{\red}{\protect\textcolor{red}}

\newcommand{\sint}{\mathop{\mathpalette\dosint\relax}\!\int}
\newcommand{\dosint}[2]{%
  \ifx#1\displaystyle
    \displaysint
  \else
    \normalsint{#1}%
  \fi
}
\newcommand{\displaysint}{\displaystyle\mathsf{s}\mkern-18mu}
\newcommand{\normalsint}[1]{%
  \smallers{#1}\ifx#1\textstyle\mkern-9mu\else\mkern-8.2mu\fi
}
\newcommand{\smallers}[1]{%
  \vcenter{\hbox{$\ifx#1\textstyle\scriptstyle\else\scriptscriptstyle\fi\mathsf{s}$}}%
}

\graphicspath{{./figures/}}

\let\OLDthebibliography\thebibliography
\renewcommand\thebibliography[1]{
  \OLDthebibliography{#1}
  \setlength{\parskip}{0pt}
  \setlength{\itemsep}{0pt plus 0.3ex}
}

\begin{document}
\theoremstyle{definition}

\newtheorem{theorem}{Theorem}

\theoremstyle{definition}
\newtheorem{definition}{Definition}

\theoremstyle{definition}
\newtheorem{corollary}{Corollary}

\theoremstyle{definition}
\newtheorem{proposition}{Proposition}

\theoremstyle{definition}
\newtheorem{lemma}{Lemma}

\theoremstyle{definition}
\newtheorem{claim}{Claim}

\theoremstyle{definition}
\newtheorem{conjecture}{Conjecture}

\theoremstyle{definition}
\newtheorem{observation}{Observation}

% \setcounter{page}{1}
% \pagenumbering{arabic}
\title {
Overcoming High Frequency Limitations of \\
Current-Mode Control Using a Control Conditioning \\  Approach --- Part II: Implementation and Hardware}
\date{}
\author{Xiaofan~Cui,~\IEEEmembership{Student Member,~IEEE,}
        and~Al-Thaddeus~Avestruz,~\IEEEmembership{Senior Member,~IEEE}% <-this % stops a space
\thanks{The authors are with the Department
of Electrical and Computer Engineering, University of Michigan, Ann Arbor,
MI, 48109 USA (e-mail: cuixf@umich.edu; avestruz@umich.edu).}% <-this % stops a space
\thanks{}}

% The paper headers
\markboth{ \hspace{1in}  {\bf DRAFT PREPRINT}}%
{Shell \MakeLowercase{\textit{et al.}}: Bare Demo of IEEEtran.cls for IEEE Journals}
\maketitle
\thispagestyle{headings}
\pagestyle{headings}

\renewcommand{\figurename}{Fig.}
\begin{abstract}
%\red{Current-mode control is one of the most popular controller strategies for power converters.}

This article is the second part of a paper series about interference in extremum (i.e.,  peak or valley) current\nobreakdash-mode control, which applies to both fixed and variable switching frequency power converters.  Specifically, this part presents three control conditioning methods that mitigate the adverse effect of interference.  These methods are new ways to use: (i) slope compensation; (ii) low-pass filtering; and (iii) the phenomenon of comparator-overdrive-delay, for control conditioning.  The stability criterion, closed-loop dynamics, and transient performance are derived with mathematical rigor for each method. 
The design tradeoffs are illustrated, discussed, and compared. 
%For the practicing engineer, a \red{we decide the the only design guide we do is for the slope compensation....} 
The effectiveness of all three methods are demonstrated and validated in hardware using a power converter operating at multi-MHz switching frequencies.

\end{abstract}

\section{Introduction}\label{sec:Intro}
\IEEEPARstart{R}{ealizing} the ultimate potential of current-mode control to be faster, more flexible, reliable, and safer is fundamentally curtailed by unwanted signals on the current sensor.  A prominent class of current-mode controller prescribes the extremum (i.e., peak or valley) of the inductor current trajectory at every switching cycle.  Both fixed frequency \cite{Yan2021a, Bao2021} and variable frequency (i.e., constant off-time \cite{Yan2021} and constant on-time \cite{Yan2021, Zhang2021}) varieties are employed, enabling cycle-by-cycle control of the inductor current.  The sensing of the current extremum requires a single-point measurement, which is especially vulnerable when the switching frequency approaches the frequency band of the interference.  Interference can lead to instabilities including those that manifest as subharmonics of the equilibrium switching frequency \cite{Cui2018a}.  The modeling of this interference within a control conditioning framework and the effect on the dynamics were logically delineated and rigorously derived in theory in Part~I of this paper.

Control conditioning approaches the repair of corruption from interference in the model of the current control loop.  The model of the current control loop consists of the static and dynamic mappings; the control conditioning methods described in this part of this paper repairs one or both of these mappings. The goals of the control conditioning methods are to: (i)~guarantee stability; (ii)~optimize control performance; (iii)~ease hardware implementation; (iv)~enable circuit integration; (v)~ease controller design; and (vi)~provide provable guarantees.

Among the control conditioning methods: (i)~first\nobreakdash-event\nobreakdash-triggering with latching, (ii)~slope compensation, (iii)~low\nobreakdash-pass filter conditioning, and (iv)~comparator-overdrive-delay conditioning have been investigated and can be deployed.
The principles and example design of (i) are illustrated in the Part I article.  In this article, the principle and design of (ii)-(iv) will be elaborated.

In repairing the dynamic mapping, (ii)-(iv) contribute additional dynamics to the current control loop.  We rigorously derive the stability criterion, closed\nobreakdash-loop dynamics, and transient performance of the current control loop for each. The closed-loop dynamics (e.g., poles and zeros) are needed particularly when the current controller is enclosed by an outer loop such as a voltage control loop.  The transient performance consists of settling and overshoot; these two metrics are most often used to compare different types of controllers, and in the context of this paper, control conditioning methods.  

Additionally, we illustrate the design tradeoffs of each of these methods through mathematically proven stability criteria and analytical expressions for transient performance.  We examine a unified framework for fairly comparing and optimally selecting among the conditioning methods to maximize the control performance for the practicing engineer. Each control conditioning method (ii)-(iv) together with hardware demonstration and validation using a dc\nobreakdash-dc converter switching at multi\nobreakdash-megahertz frequencies is provided. Straightforward hardware design was possible with a hybrid digital/analog strategy.

In the the Part I article, we develop the foundations for the mathematical modeling of interference as an uncertainty in the model of the plant for the current control loop and the theoretical framework for control conditioning as a mitigation of the deleterious effects.  The important concepts of static and dynamic mappings of the current control loop are discussed and how the corruption of these mappings by interference leads to subharmonic instabilities, which make the output ripple hard to filter reliably.

In this Part II article, a thorough discussion of the theory and hardware results are presented for each of the (ii)-(iv) control conditioning methods.
In Section\,\ref{sec:solution}, the large-signal stability criteria, settling, and overshoot are theoretically derived in closed-form for each of the methods, with proofs provided in papers \cite{Avestruz2022a, Avestruz2022b}. In addition, intuitive graphical and textual descriptions of the dynamics are included.  %In Section III, we present \red{design guides} for the practicing engineer and use hardware results to illustrate as well as validate each of the control conditioning methods.  
%Within each \red{design guide}, the tradeoffs curves between settling and overshoot are provided. 
For each method, the hybrid analog/digital hardware implementation is carefully described in Section \ref{sec:hardwaredesignguide}. 
In Section \ref{sec:conlusion_part2}, a summary of the results and contributions of this paper are summarized.
\begin{figure}
    \centering
    \includegraphics[width=8cm]{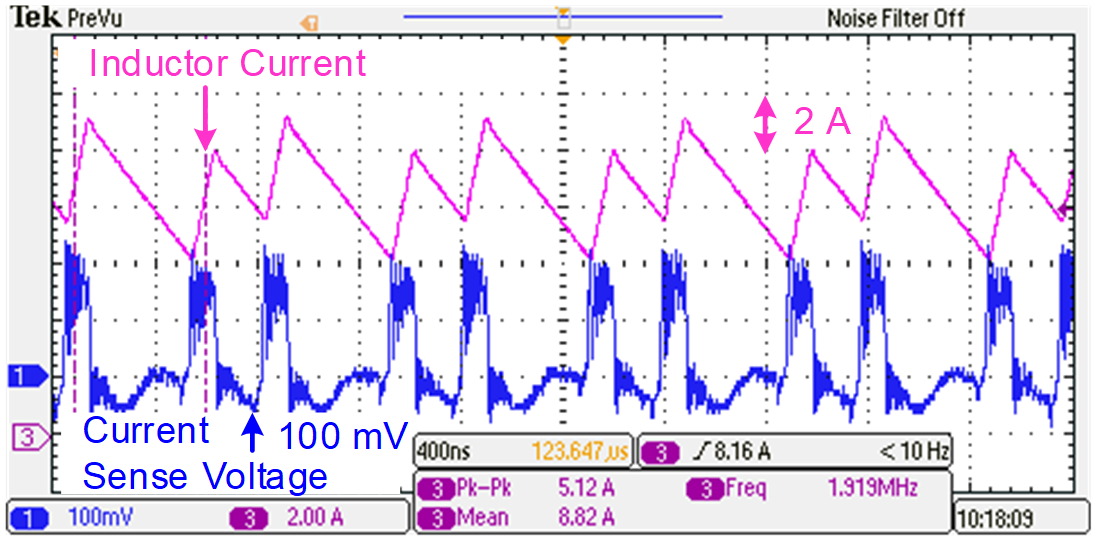}
    \caption{Current sense voltage of a current\nobreakdash-mode buck converter using constant on\nobreakdash-time (10\,m$\Omega$ current sense resistor). The current sensor output is largely distorted by interference, and the measurement error can be as much as $40\%$. The interference comes from the parasitic ringing and power ground resonance. Subharmonics on the inductor current waveform manifest because the interference severely destabilizes the current control loop.} \label{fig:noisecurrent_part2}
\end{figure}

\section{Design of Control Conditioning Methods} \label{sec:solution}
Control conditioning changes both the static and dynamical mappings of the current control loop to flatten nonlinearities, stabilize, and improve transient performance from what had been degraded by interference.  For the static mapping, control conditioning transforms the interference to reduce the degree of nonlinearity.  One unintended consequence of this interference transformation is the transformation of the ideal static mapping.  In the best case, the ideal static mapping remains a line, but may be offset or changed in slope; however, this transformation can additionally introduce nonlinearities to this ideal mapping.  The best-fit line to this transformed ideal mapping can be considered the ``new ideal mapping'', which becomes the baseline for the analysis of deviations discussed in Section\,
II of the Part I article \cite{cmpartone2022}. From this analysis, the deviation results from (a) the unintended nonlinearity from control conditioning and (b) the nonlinearity from interference.

For dynamical mapping, control conditioning transforms the interference to reduce the worst-case model gains that cause instability, slow settling, and large overshoot. One unintended consequence of this interference transformation may be the introduction of additional destabilizing gains, poles, zeros, or a delay, which degrades the stability margin. Designing the control conditioning involves balancing two competing mechanisms: the reduction of the deleterious effects of interference and the unintended dynamics introduced by control conditioning.

We discuss four new control conditioning methods to confront interference.  First-event triggering with latching is the precursor to the following three.  The multivalued static mapping can be resolved by first-event triggering with latching.
However, this triggering criterion can cause discontinuities in the static mapping.  These discontinuities along with other nonlinearities can be repaired by the following three methods: slope compensation, low-pass filtering, and comparator overdrive delay. These methods also affect the dynamical mapping whose impact to stability must be considered, leading to different tradeoffs in the transient performance. Our theoretical results enable the guarantee of {\em global asymptotic stability} of the dynamical mapping and hence the current control loop, while optimizing the transient performance.  The low-pass filter, although the most often-used method for alleviating interference, is typically selected in an ad-hoc manner.  The filter is the most straightforward to implement in hardware; however, guarantees of stability are theoretically involved and usually result in worse transient performance than the other two methods, and are often infeasible when the band of interference is lower or near the switching frequency.

The other two methods come from two familiar constituents of power converters, but are used in an entirely new way.  Slope compensation is the most straightforward to understand as a control conditioning method.  Although traditionally used to stabilize a different phenomenon \cite{Redl1981a}, using this well-known method to alleviate interference leads to a surprising result. The overdrive delay in a comparator as a means for control conditioning is original.  This method can readily be implemented as part of an integrated circuit controller and the delay can easily be made tunable.  

\begin{figure}
    \centering
    \includegraphics{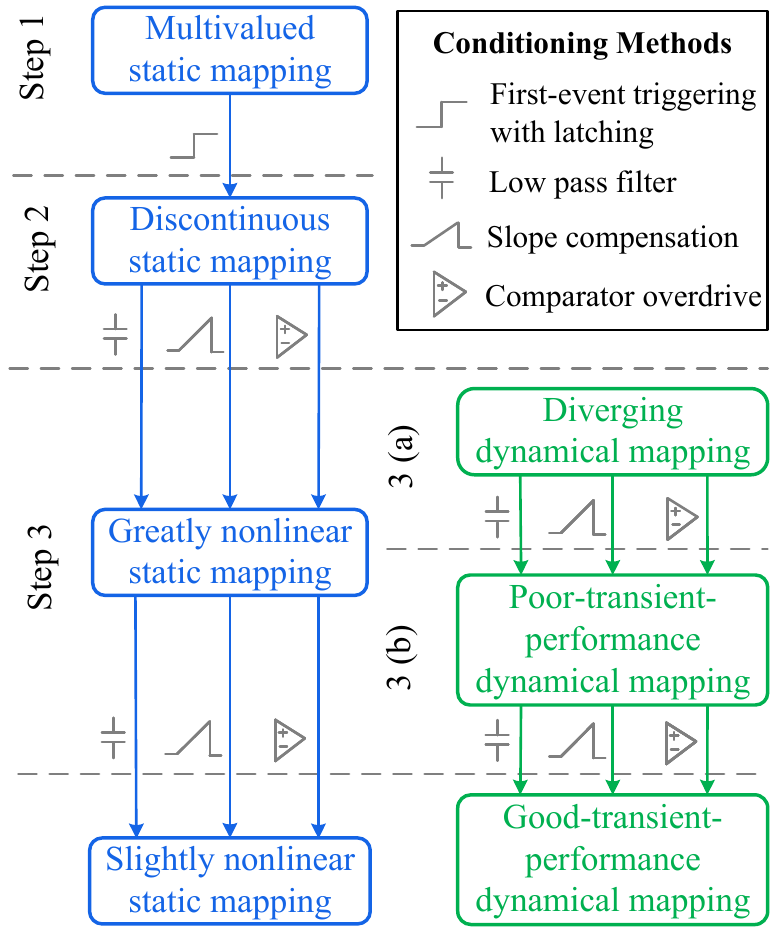}
    \caption{3-steps design procedure of control conditioning. Four conditioning methods --- first-event triggering with latching, low-pass filtering, slope compensation and comparator overdrive are applied to repair the defective static mapping and dynamical mapping step-by-step.}
    \label{fig:designflow_cond}
\end{figure}

In this section, we outline and rigorously prove a 3-step design procedure to precisely design the control conditioning of interference in current-mode control.
We offer quantitative comparisons in choosing slope compensation, low pass filtering, and comparator overdrive.
\begin{enumerate}
    \item In step 1, we use first-event triggering with latching to resolve the problem of the multivalued static mapping. 
% The ultimate goal is to decrease the nonlinearity of the static mapping, and make dynamical mapping stable, shorter settling time and smaller overshoot in transition.
% We explore four new conditioning methods.
    \item In step 2,
% we show three different conditioning methods to repair the discontinuous static mapping.
we use one of three conditioning methods including slope compensation, low-pass filtering, and comparator overdrive delay to condition the discontinuous static mapping.
Each of these can independently solve the problem.
We prove the condition for each method to make the static mapping continuous.

\item In step 3, 
% these three conditioning methods can also be used to decrease the degree of nonlinearity of the static mapping.
we carefully design these three methods to condition the degraded dynamical mapping. We prove the criterion for each method to make the dynamical mapping stable.
% under the premise of stability, we carefully design three methods to optimize the transient performance. We mainly focus on two performance criteria --- settling time and overshoot. 
we analyze the design trade-off for each method to optimize the settling and overshoot of the dynamical mapping.
The design procedure is summarized in Fig.\,\ref{fig:designflow_cond}.
\end{enumerate}

\subsection{Slope Compensation} \label{sec:solution_subsec:slopecompensation}
\begin{figure}
    \centering
    \includegraphics[width = 8cm]{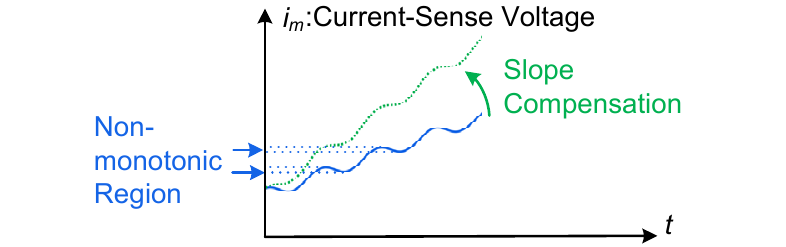}
    \caption{Time domain explanation of how slope compensation conditions the interference. Slope compensation guarantees the continuity of the static current mapping by rotating the current sensor signal until it becomes monotonic and re-scaling the deviation from the interference.}
    \label{fig:slope_w_inf_td}
\end{figure}

\begin{figure}
    \centering
    \includegraphics{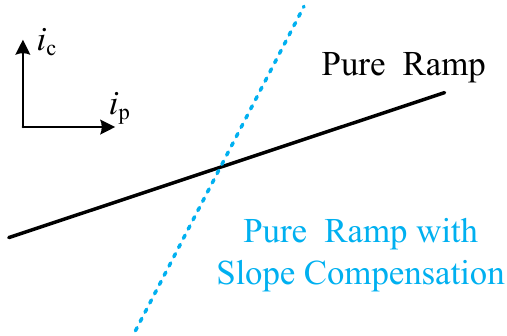}
    \caption{Slope compensation does not affect the nonlinearity of the static current mapping but causes a gain error on the mapping.}
    \label{fig:slope_no_inf}
\end{figure}

Slope compensation is a well-recognized control conditioning method that stabilizes the fixed-frequency current-mode control loop when the duty cycle crosses 50\,\%.
What had been previously unknown is that slope compensation can also be used for the control conditioning of interference.
% We explore the new applications of slope compensation for interference conditioning in both constant on-time and fixed-frequency current-mode controller.
Slope compensation can repair both discontinuities and the ensuing smooth nonlinearities in the static mapping, as well as improve the transient performance of the dynamical mapping.
Slope compensation decreases the degree of nonlinearity by effectively flattening the interference in the time domain waveform as shown in Fig.\,\ref{fig:slope_w_inf_td}. This has the consequence of flattening the static mapping by contracting the points in the interference that correspond to early triggering (negative deviation from transformed ideal line of the current mapping) and late triggering (positive deviation from the transformed ideal line) as illustrated in Fig.\,\ref{fig:slope_w_inf_td}.  For an ideal static mapping, slope compensation does not introduce nonlinearities, but rather introduces a slope (gain) error in the mapping, as shown in Fig.\,\ref{fig:slope_no_inf}.  As previously mentioned, this gain error is outside of the current control loop and is subsequently corrected in the design of the outer voltage loop.
% The extra negative slope is equivalently an extra positive slope on the current ramp. It improves the signal-to-interference ratio of the sensed inductor current by enhancing the signal as shown in Fig.\;\ref{fig:sc1}. Hence the stability of current-controller is improved.
% \begin{figure}
%     \centering
%     \includegraphics[width=8cm]{Figure/section3/slopecompensation/slopecomp.pdf}
%     \caption{ \label{fig:sc1} Comparison of the current-sense voltage with/without slope compensation. {\color{blue}---} is for the current-sense voltage without slope compensation. {\color{orange}$\cdots$} is for the current-sense voltage with slope compensation. {\color{red}-\,-\,-} is for the compensation slope.}
% \end{figure}
\begin{figure}
    \centering
    \includegraphics[width = 6cm]{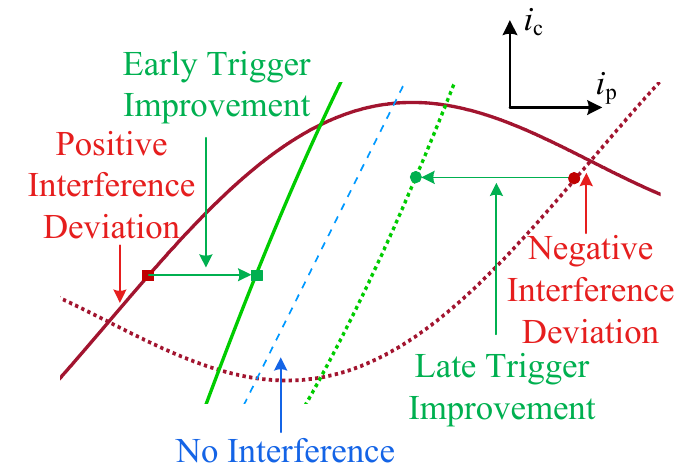}
    \caption{Slope compensation weakens the nonlinearity of the static current mapping by contracting the deviation from the ideal triggering.}
    \label{fig:slope_w_inf}
\end{figure}

% \subsubsection{Modeling of the current-controller with Slope Compensation}
We start with constant off-time current-mode control as an example.
By adding slope compensation, we replace $i_c$, which is constant within every switching cycle, with \mbox{$i_c - m_s t$}, which is a decreasing ramp for $m_s>0$.
A state-space form of the current control loop is
%\begin{subequations}
    \begin{align} \label{eqn:slopecompsys}
     %\label{eqn:slopecompsysip} 
     i_p[n] &= i_p[n-1] - m_2T_{\text{off}} +m_1t_{\text{on}}[n], \nonumber \\
     i_c [n]  &= i_p[n] + w(t_{\text{on}}[n]) + m_s t_{\text{on}}[n].
    \end{align}
%\end{subequations}

%\begin{subequations}
%\label{eq:Maxwell}
%Maxwell's equations:
%\begin{align}
%        B'&=-\nabla \times E,         \label{eq:MaxB} \\
%        E'&=\nabla \times B - 4\pi j, \label{eq:MaxE}
%\end{align}
%\end{subequations}

% \subsubsection{Stability Analysis}
% The Proposition\,\ref{th:equexcond} indicates that the static current mapping is continuous if and only if the inference does not affect the monotonicity of the current sensor output.

Proposition\,1 in the Part I article \cite{cmpartone2022} states that if the current sensor output is not monotonic, then the static current mapping is discontinuous.
A discontinuous static mapping can make the current control loop unstable.
Corollary\,\ref{reachthoerem:slopecomp_constontime} below states that if a slope compensation $m_s$ is added, the discontinuity can be eliminated as had been discussed for Fig.\,\ref{fig:slope_w_inf_td}.
We show that an appropriately designed slope compensation can make the time\nobreakdash-domain function of the current sensor output monotonic;
from Proposition\,1 in the Part I article  \cite{cmpartone2022}, the static mapping is also made continuous.
\begin{corollary} \label{reachthoerem:slopecomp_constontime}
$\mathcal{T}$ is a strictly monotonically increasing and continuous static mapping if and only if  \mbox{$(m_1 + m_s)\,t+w(t)$} is strictly monotonically increasing and continuous within a switching cycle.
\end{corollary}
% The proof relies on the similarity of system
% (\ref{eqn:slopecompsys}) with (\ref{eqn:cmcotshiftlin}), hence it is omitted in this paper.

In this section, we show that adding an extra slope transforms both the interference and the ideal current ramp, which allows the dynamical mapping to be stabilized even if the original interference violates the stability bound in Theorem\,1 of the Part I article \cite{cmpartone2022}. From Theorem\,1 of the Part I article \cite{cmpartone2022} and having satisfied its condition of continuous static mapping, the stability of the dynamical mapping depends on the upper bound of the Lipschitz constant $\Lambda_{ub}$ of the interference.  For a larger $\Lambda_{ub}$, an accordingly larger $m_s$ is needed to make the current control loop globally asymptotically stable.

This slope compensation method does not impose any restrictions on the amplitude and frequency of the interference, which means it can be an often-used and robust large-signal method to stabilize the current control loop.  One proviso is that a larger compensation $m_s$ corresponds to a higher loop gain, hence moving the pole on the $z$-domain root locus as we explain below; when it is too high, it slows down the closed\nobreakdash-loop response of the current control loop, which is the opposite of what happens in the $s$-domain when the loop gain becomes higher.

%{\color{red} For the same amplitude, higher frequency interference %= higher Lipschitz constant.
%So need higher slope compensation.
%Higher slope compensation means higher loop gain, which moves the %pole from the origin towards the unit circle -->  slowing down the %loop bandwidth.
%}

\begin{figure}[htp]
    \centering
    \includegraphics[width=9cm]{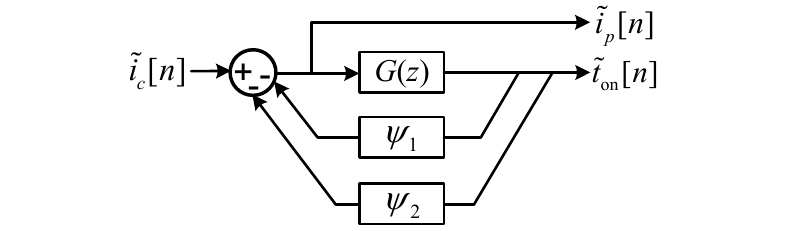}
    \caption{Large-signal block diagram of the constant off-time/fixed-frequency peak current control loop with slope compensation. The interference is embedded in $\psi_1$ and the slope compensation is embedded in $\psi_2$.}
    \label{fig:sccotcmblockdiagram}
\end{figure}

We approach the stability analysis by applying the $z$\nobreakdash-transform to system (\ref{eqn:slopecompsys}).
The current control loop can be represented as the Lure system in Fig.\;\ref{fig:sccotcmblockdiagram} with
\begin{align} \label{eqn:gzofcotcm_part2_sc}
G(z) = \frac{1-z^{-1}}{m_1}.
\end{align}
%(\ref{eqn:cmcotip_sc}) and %(\ref{eqn:cmcotic_sc}) 
% \begin{align}
%     G(z) = \frac{1-z^{-1}}{m_1}.
% \end{align}
The interference is embedded in the static nonlinearity $\psi_1$ in the feedback path
\begin{align}
    \psi_1 = w(\tilde{t}_{\text{on}}[n] + T_{\text{on}}[n]) - w( T_{\text{on}}[n]).
\end{align}
The slope compensation is embedded as a gain $\psi_2$ also in the feedback path 
\begin{align}
    \psi_2 = m_s.
\end{align}
% \begin{align} \label{eqn:S2_2}
% s = \frac{f^{'}(T_{\text{on}}) + m_s}{m_1}.
% \end{align}
If the function $w(\cdot)$ is monotonically decreasing, 
the feedback path $\psi_1$ is a positive feedback path and might cause instability.
Typically the compensation slope $m_s$ is chosen to be positive, which makes $\psi_2$ a negative feedback path, which can be used to correct the destabilizing effect of $\psi_1$.

As a Lure system, we can apply the circle criterion similarly to the Part I article; the large-signal stability criterion for the dynamical mapping can be proven.
\begin{corollary} \label{stabthoerem:slopecomp_constontime}
The current control loop represented by the Lure system in Fig.\,\ref{fig:sccotcmblockdiagram} is \emph{globally asymptotically stable} if \mbox{${\Lambda}_{ub} < m_1/2 + m_s$}.
\end{corollary}
% The proof is similar to  Theorem\,\ref{theorem:gloasystab}, hence omitted in this paper.
% \begin{figure}
% ~
% \begin{minipage}{0.5\textwidth}
%     \centering
%     \includegraphics[width=\textwidth]{Figure/section3/slopecompensation/sc_cotcm.pdf}
%   \caption{\label{fig:sc2} Root locus of the current-controller of constant on(off)-time control without(left)/with(right) the slope compensation. The solid green line is for the negative feedback and the dotted blue line is for the negative feedback.}
% \end{minipage}
% % ~
% % \begin{minipage}{0.32\textwidth}
% %     \centering
% %     \includegraphics[width=\textwidth]{Figure/section3/copd/physicalmodelcomp.PNG}
% %     \caption{\label{fig:physicalmodel} Simplified physical model of a comparator.}
% % \end{minipage}
% \end{figure}

% we start the subfig here
\begin{figure}
    \centering
    \includegraphics[width= 8 cm]{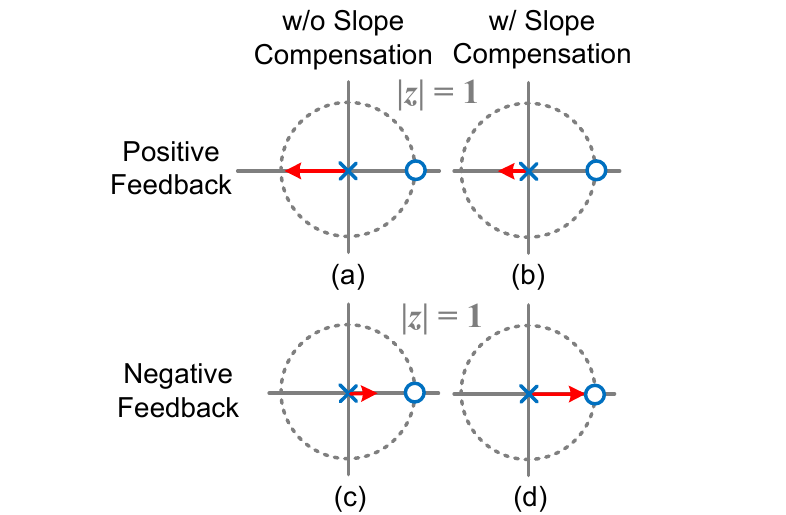}
    \caption{Small-signal root locus of the constant off(on)-time current control loop with the slope compensation. The sign of the feedback is determined by the interference.  In positive feedback, slope compensation decreases the loop gain to increase the stability margin.  In negative feedback, the converter is always stable, but results in a slower transient performance. In practice, the polarity of the feedback cannot be determined a-priori and the compensation slope is chosen so that the converter is stable in positive feedback.}
    \label{fig:sc_cotcm_new}
\end{figure}
We can obtain an intuition for stability by examining the linearized system of (\ref{eqn:slopecompsys}).
% \begin{align} 
% \label{eqn:cmcotip_sc} \tilde i_p[n] &= \tilde i_p[n-1] +  m_1 \tilde t_{\text{on}}[n], \\
% \label{eqn:cmcotic_sc} \tilde i_c [n]  & = \tilde i_p[n] + (f^{'}(T_{\text{on}}) +m_s)\tilde t_{\text{on}} [n].
% \end{align}
% We normalize the slope by defining the slope ratio $\gamma = m_s/m_1$ and $s = f^{'}(T_{\text{on}})/m_1$. 
The root locus for this linearized system is illustrated in Fig.\,\ref{fig:sc_cotcm_new} and helps to visualize how slope compensation improves stability and affects the transient response. 
% The slope compensation right shifts the poles which were outside of the unit-cycle into the unit-cycle.
% Slope compensation can make unstable poles into stable poles; it does this by right shifting the poles into the unit circle.
% Performance analysis, by doing linearization and Z-transform
% The small-signal model is:
% \begin{align}
% \tilde i_p[n] &= \tilde i_p[n-1] +  m_1 \tilde t_{\text{on}}[n], \\
% \tilde i_c [n] - m_s \tilde t_{\text{on}} [n] & = \tilde i_p[n] + f(t_{\text{on}}[n]) - f(T_{\text{on}}).
% \end{align}
% Then the new nonlinear static function is $\psi(y) = m_1 \phi ^{-1}(y-f(T_{\text{on}}))$, where $\phi(x) = m_s x + f(x)$. If (\ref{sc1}) holds, we can confidently say the oscillation cannot sustain in the inner loop. 
% \begin{align} \label{sc1}
% m_s + (f')_{\text{min}} > -m_1/2
% \end{align}
% \subsubsection{Transient Performance Analysis}
% {\color{red} (revised by XC for review) If the interference causes a positive feedback path, the closed-loop pole locates at the left half plane and may be outside of the unit circle as shown in Fig.\,\ref{fig:sc2}(a).
% Adding the slope compensation can right shift the unstable pole into the unit cycle so that the current control loop is stabilized as shown in Fig,\,(b).
% If the interference causes a negative feedback path, the closed-loop pole locates at the right half plane as shown in Fig.\,\ref{fig:sc2}(c).
% Adding the slope compensation may right shift the closed-loop pole towards $z = 1$ and increase the worst-case settling.
% }
We observe that slope compensation for positive feedback in Figs.\,\ref{fig:sc_cotcm_new} (a) and (b) moves the location of the worst\nobreakdash-case closed-loop pole further to the right, hence improving stability margin.  Likewise, for negative feedback in Figs.\,\ref{fig:sc_cotcm_new} (c) and (d), adding the slope compensation right\nobreakdash-translates the closed\nobreakdash-loop pole towards \mbox{$z = 1$} and increases the worst-case settling.
% So there exists a $\tau_{\text{min}}$ such that for all $\tau > \tau_{\text{min}}$, all the poles of the current-command block can stay inside the unit-cycle. The following Theorem provides an accurate design method for the filter.

We use the settling and overshoot metrics in (\ref{eqn:settlecycle1_part2}) and (\ref{eqn:overshoot1_part2}), respectively, to quantitatively illustrate the relationship between these metrics and the pole locations. With interference, the pole $a$ cannot be located exactly, but rather within a range $[a_{\text{min}},a_{\text{max}}]$, where
\begin{align} %\label{eqn:range_pole_sc}
a_{\text{min}} = \frac{m_s-{\Lambda}_{ub}}{(m_1 + m_s - {\Lambda}_{ub})},  
\,a_{\text{max}} = \frac{m_s+{\Lambda}_{ub}}{(m_1 + m_s + {\Lambda}_{ub})}.
\end{align}
The worst-case for settling and overshoot can be obtained from (\ref{eqn:settlecycle1_part2}) and (\ref{eqn:overshoot1_part2}), respectively. It is worth noting that $a_{\text{min}}$ and $a_{\text{max}}$ are not usually symmetric about the origin.

\begin{figure}[htp]
\centering
\subfigure[The worst-case settling $N_w$ decreases first and then increases with the compensation slope $\hat{m}_s$.]{
\includegraphics[width=6.5 cm]{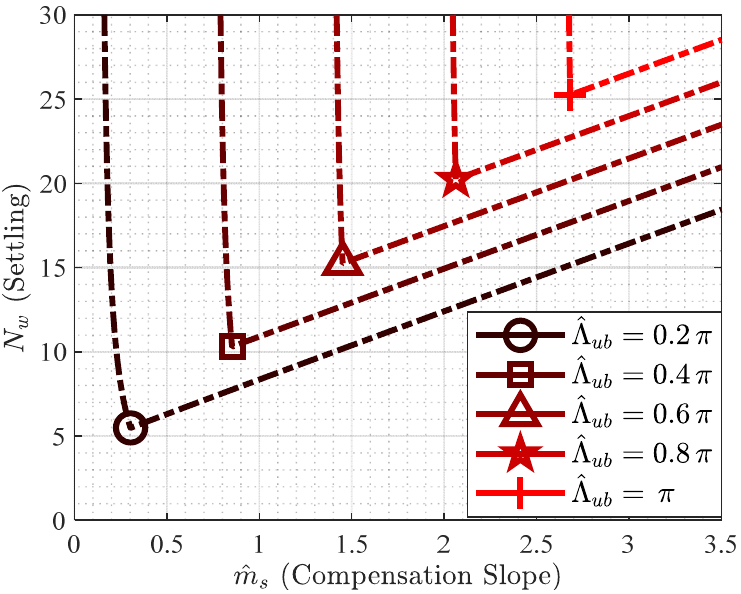} \label{fig:scsettlingtime}
}
\subfigure[The worst-case overshoot $O_w$ decreases with the compensation slope $\hat{m}_s$.]{
    \includegraphics[width=6.5 cm]{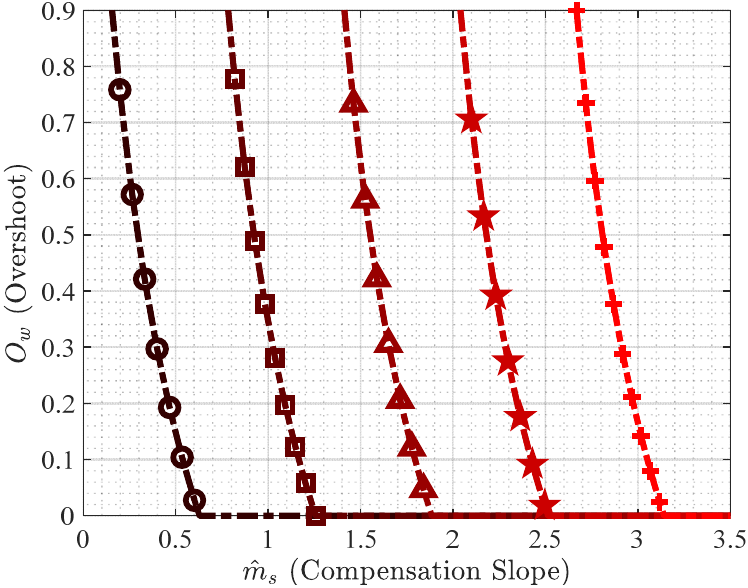}  \label{fig:scovershoot}
}
\caption{The control conditioning diagram for compensation slopes given different interference levels $\hat{\Lambda}_{ub}$. For a given interference, there is a minimum settling $N_w$ required. Beyond a certain compensation slope, the overshoot is always zero, but the penalty in $N_w$ keeps increasing.}
% Caption for if you have a tradeoff curve:  For a given interference, there is a minimum settling $N_w$ required and beyond a certain compensation slope, the overshoot is always zero, but the penalty in $N_w$ keeps increasing; this is exemplified in the tradeoff curve.
%\label{fig:subfigureExample}
\end{figure}

The transient performance of the current control loop varies with the compensation slope, as shown in Fig.\,\ref{fig:scsettlingtime} and Fig.\,\ref{fig:scovershoot}.
The transient performance is related to the \emph{normalized} compensation slope $\hat{m}_s$ and \emph{normalized} Lipschitz constant of interference $\hat{\Lambda}_{ub}$
\begin{align} \label{eqn:ms_norm}
    \hat{m}_s \triangleq \frac{m_s}{m_1},\quad \quad \hat{\Lambda}_{ub} \triangleq \frac{{\Lambda}_{ub}}{m_1}.
\end{align}

\begin{figure}
    \centering
    \includegraphics[width = 8cm]{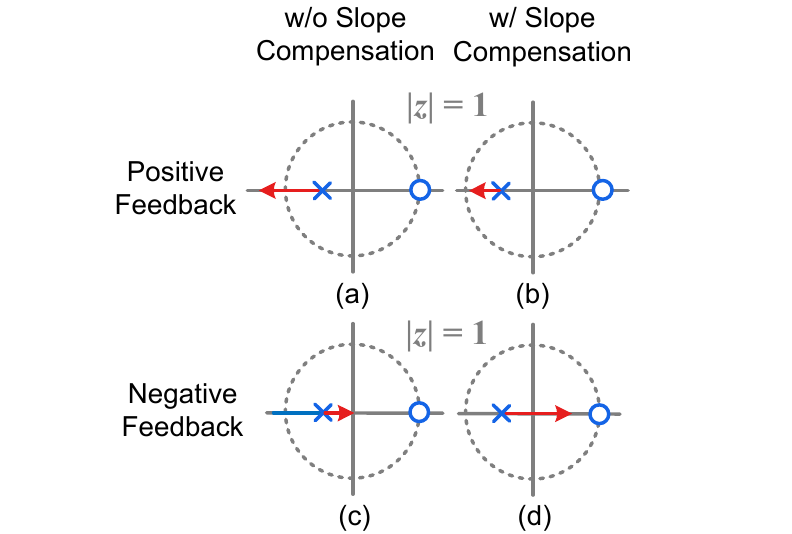}
    \caption{Small-signal root locus for fixed-frequency current control loop with slope compensation. 
    The sign of the feedback is determined by the interference.  In positive feedback, slope compensation decreases the loop gain to increase the stability margin.  In negative feedback, the converter is always stable, but results in a slower transient performance. In practice, the polarity of the feedback cannot be determined a-priori and the compensation slope is chosen so that the converter is stable in both positive and negative feedback.}
    \label{fig:rl_fpcc_sc}
\end{figure}

The curves from Fig.\,\ref{fig:scsettlingtime} are convex with points of minimum settling, which are marked by symbol points on each of the curves. The worst-case overshoot decreases monotonically with the compensation slope as shown in Fig.\,\ref{fig:scovershoot}. These show a tradeoff between settling and overshoot.
%Therefore, given any design objective which can be formulated as a combination of settling time and overshoot, the optimal slope can be found by the control conditioning diagram. 
The range of pole locations can be expressed as
\begin{align} %\label{eqn:range_pole_sc}
a_{\text{min}} = \frac{\hat{m}_s-\hat{\Lambda}_{ub}}{(1 + \hat{m}_s - \hat{\Lambda}_{ub})},\,\, a_{\text{max}} & = \frac{\hat{m}_s + \hat{\Lambda}_{ub}}{(1 + \hat{m}_s + \hat{\Lambda}_{ub})}.
\end{align}
% The compensation $\hat{m}_s$ over different degrees of interference $\hat{\Lambda}_{ub}$ are shown in Fig.\,\ref{fig:scsettlingtime}.
By observing that the bounds $a_{\text{min}}$ and  $a_{\text{max}}$ monotonically move to the right in the root locus with increasing $\hat{m}_s$, the criterion for minimum worst-case settling is then $a_{\text{min}} = -a_{\text{max}}$, from which the corresponding $\hat{m}_s^*$ can be solved
\begin{align} \label{eqn:optms}
     \hat{m}_s^* & = \sqrt{\frac{1}{4} + {\hat{\Lambda}_{ub}}^2} - \frac{1}{2}.
\end{align}
% The detailed proof can be found in Appendix  \ref{sec:proofoptsc}.
The corresponding minimum worst-case settling is
\begin{align} \label{eqn:th_wc_settling}
    N_w^* = 
    \Bigg|\frac{4}{\text{ln}\Big|1 - (1 + \sqrt{\frac{1}{4}+{\hat{\Lambda}_{ub}}^2} + \hat{\Lambda}_{ub})^{-1}\Big|}\Bigg|.
\end{align}
% The effect of poles by the ramp compensation
% Formulate an optimization
% \subsubsection{Analysis on Constant On-time and Fixed-frequency current-controller}
These results can be extended to other types of current control loops.
The continuity theorem, stability theorem, and the performance analysis for constant on-time control can be derived by replacing $G(z)$ by
\begin{align}
       G(z) = \frac{(z^{-1} - 1)}{m_2}.
\end{align}
The result for fixed-frequency control can be derived by replacing $G(z)$ by  
\begin{align}
        G(z) &= \frac{1-z^{-1}}{m_1 + m_2 z^{-1}}.
\end{align}
The effect of slope \mbox{compensation} on the root locus for fixed-frequency control is shown in Fig.\;\ref{fig:rl_fpcc_sc}.
\subsection{Low-Pass Filter Conditioning} \label{sec:solution_subsec:lowpassfilter}
Traditionally, the low\nobreakdash-pass filter is used for signal conditioning to \mbox{reduce} the amplitude of the interference signal when the interference frequency is well above the switching frequency.
In this way, the cut\nobreakdash-off frequency of the low\nobreakdash-pass filter can be above the switching frequency hence attenuating the amplitude of the
interference without affecting the inductor current ramp.
% From the perspective of control conditioning, the low\nobreakdash-pass filter not only reduces the amplitude of the interference signal but also warps the static mapping and introduces a new smooth nonlinearity.
% Therefore, The cut\nobreakdash-off frequency of the low\nobreakdash-pass filter can be designed to be nearly at or below the switching frequency.
When the time scale of the interference is near or below the switching frequency, the cut\nobreakdash-off frequency will also be near or below the switching frequency; through control conditioning, the introduction of a now significant and undesirable smooth nonlinearity is considered. In doing so, stability and good transient performance can be guaranteed.

% The difference is control conditioning approach treats the warping effect of ramp intrinsic while signal conditioning approach treats the warping effect of ramp extrinsic.
% Because of this, the control conditioning approach overcomes the limitation that low\nobreakdash-pass filter can only deal with the interference whose spectrum is around or below the switching frequency.
% From the control conditioning perspective, you can actually account for the fact that the cut\nobreakdash-off frequency is nearly at or below the switching the frequency.

Low\nobreakdash-pass filtering can repair both the discontinuity and nonlinearity problem in the static mapping.
% Filter can also improve the transient performance of dynamical mapping.
Despite the introduction of an unwanted nonlinearity, the filter repairs discontinuities in the static mapping, and can actually improve the interference\nobreakdash-degraded transient performance of the dynamical mapping by reducing the interference amplitude.

\begin{figure}
    \centering
    \includegraphics[width = 6cm]{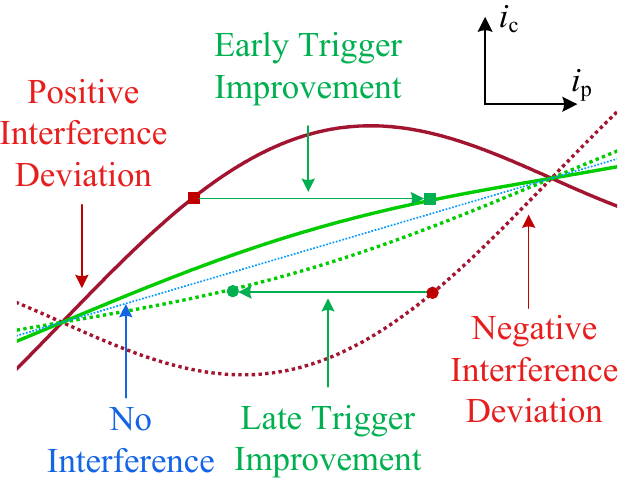}
    \caption{Filter shows a weakening effect on the nonlinearity of the static current mapping because it attenuates the interference signal.}
    \label{fig:filter_imap_with_inf}
\end{figure}

\begin{figure}[htp]
    \centering
    \includegraphics{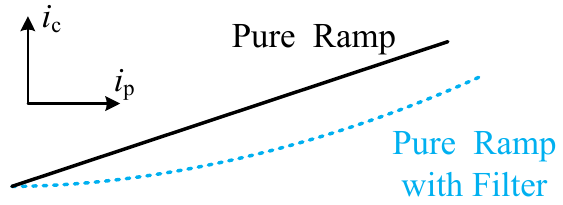}
    \caption{Filter shows an enhancing effect on the nonlinearity of the static current mapping because it warps the ramp signal.}
    \label{fig:filter_imap_no_inf}
\end{figure}

The filter decreases the degree of nonlinearity in the static mapping, as shown in Fig.\,\ref{fig:filter_imap_with_inf}.
The early trigger, which is caused by a positive interference deviation, is delayed by the filter. 
The late trigger, which is caused by a negative interference deviation, is advanced by the filter.
The filter makes an ideal static mapping nonlinear as shown in Fig.\,\ref{fig:filter_imap_no_inf}, while also able to make a larger nonlinearity from interference smaller.  Therefore, there exists an optimal filter to perform the control conditioning.
% Low-pass filter, the traditional way to tackle against the interference, stabilizes the current control loop by attenuating the amplitude of interference. However, as shown in Fig.\;\ref{fig:filter1}, the actual inductor current may be distorted. In this section, we develop a theory to optimize the design of the filter.
% \subsubsection{Modeling of the current bluecontroller with low-pass filter}

The current control loop using constant off\nobreakdash-time can be modeled as
% \begin{align}
% i_p[n] &= i_p[n-1] - m_2T_{\text{off}} +m_1t_{\text{on}}[n], \nonumber \\
% i_c [n] &= (i_p[n-1] - m_2T_{\text{off}})(1-e^{-t_{\text{on}}[n]/\tau}) 
%      + \int_0^{t_{\text{on}}[n]} m_1 (1-e^{t/\tau}) dt  \nonumber \\
%      & + i_c [n-1] e^{-T_{\text{off}}/\tau} e^{-t_{\text{on}}[n]/\tau} 
%      + \frac{A \, \text{sin}(\omega  t_{\text{on}}[n]+\varphi)}{\omega \tau}.
% \end{align}
\begin{subequations}
\begin{align} 
    \label{eqn:filter_nonlin_sys} i_p[n] =\,& i_p[n-1] -m_2T_{\text{off}} +m_1\,t_{\text{on}}[n],\\
 %h(t)\bigg_{t=t_{\text{on}}[n]},
    \label{eqn:filter_nonlin_sys_fb} i_c [n] =\,& h_0(t_{\text{on}}[n]) h_0(T_{\text{off}}) i_c[n-1]
    \nonumber \\
    & + \left(i_m(t) u(t) \mathop{\scalebox{1.5}{$\ast$}} h(t)\right)\bigg\rvert_{t=t_{\text{on}}[n]},
\end{align}
\end{subequations}
where $\mathop{\scalebox{1.5}{$\ast$}}$ is the convolution operator, $h(t)$ is the impulse response of the low\nobreakdash-pass filter, $h_0(t)$ is the zero input response of the filter, and $u(t)$ is the unit step function.
%The current sensor output, which corresponds to the filter input, $i_m(t)$ can be expressed as the additive summation of inductor current on the bottom switch and interference.
%Equation (\ref{eqn:filter_nonlin_sys_fb}) captures the low\nobreakdash-pass filter response.
%$i_m(t)$ represents the current sensor output and filter input.
$i_m(t)$ can be expressed as the additive summation of the inductor current during the on time and the interference,
\begin{align}
    i_m(t) = i_p[n-1] - m_2T_{\text{off}} + m_1 t + w(t).
\end{align}
% The desired current $I_c$ and actual current $I_p$ at the equilibrium satisfy
% \begin{align} \label{eqn:equic_front}
%     I_c = \frac{\left(\left( I_p -m_2T_{\text{off}} + m_1 \, t + w(t) \right) u(t){\ast}h(t)\right)\bigg\vert_{t=T_{\text{on}}}
%     }{1-q(T_{\text{on}})q(T_{\text{off}})}.
% \end{align}
% \begin{figure}
%     \centering
%     \includegraphics[width=8cm]{Figure/section3/filter/filter.pdf}
%     \caption{Comparison of current-sense voltage with/without the interference and with/without the filter. {\color{blue}---} is for the current-sense voltage with interference and without filter. {\color{red}$\cdots$} is for the current-sense voltage with interference and with filter. {\color{green}-\,$\cdot$\,-} is for the current-sense voltage without interference and without filter. {\color{orange}-\,-\,-} is for the current-sense voltage without interference with filter.}
%     \label{fig:filter1}
% \end{figure}
\begin{figure}[htp]
\centering
\includegraphics[width=8 cm]{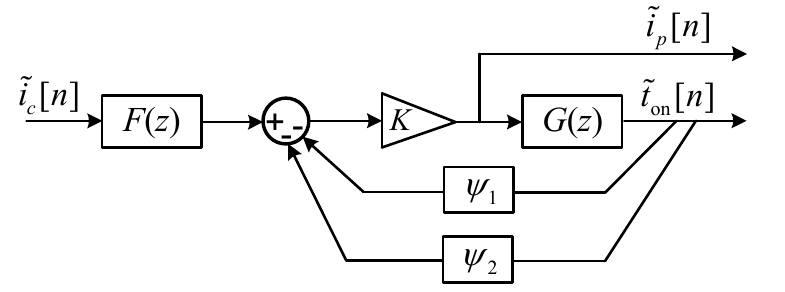}
\caption{Small-signal block diagram of the constant on-time/fixed-frequency peak current control loop with the filter.} \label{fig:filteredccb_fb}
\end{figure}

\begin{figure}[htp] 
\centering
\includegraphics[width=6 cm]{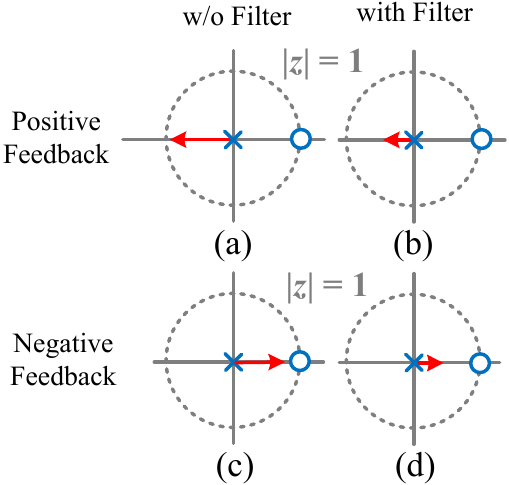}
\caption{Small-signal root locus of the constant off(on)-time current control loop with the low-pass filter. The sign of the feedback is determined by the interference. In positive feedback, low-pass filter decreases the loop gain to increase the stability margin. 
In negative feedback, the closed\nobreakdash-loop pole is moved further to the left by the filter.}
\label{fig:filter_cotcm}
\end{figure}

A first\nobreakdash-order low\nobreakdash-pass filter has the impulse response \mbox{$h(t) = u(t)\,e^{-t/\tau}/\tau$} and \mbox{$h_0(t) = e^{-t/\tau}$} where $\tau$ is the time constant.  As mentioned previously, a continuous static mapping is a prerequisite for stability.
% We assume that the cut-off frequency of filter is much smaller than the frequency of the interference function $w(t)$
% \begin{align} \label{eqn:timescaleassumption}
%     \omega_l \tau \gg 1,
% \end{align}
Theorem\,\ref{theorem:taomin4continuity} provides a sufficient condition for the filter to guarantee a continuous static mapping. As long as the time constant satisfies Theorem\,\ref{theorem:taomin4continuity}, the static mapping is continuous. We denote the lower bound of the frequency of interference by $\omega_l$.
\begin{theorem} \label{theorem:taomin4continuity}
A current control loop using constant off\nobreakdash-time has minimum on time $T^{\text{min}}_{\text{on}}$ and off time $T_{\text{off}}$. The time constant of the first\nobreakdash-order low\nobreakdash-pass filter is $\tau$. The interference $w(t)$ satisfy Definition\,2 in the Part I article \cite{cmpartone2022}.
The condition for $\tau$ to guarantee the \emph{continuous} static mapping is
\begin{align} % \label{thoerem:filterstab}
 \frac{\hat{A}_{ub}}{(1-d)\hat{\tau}}\left( 1 + \frac{d}{\sqrt{1 + (2\pi \hat{\omega}_l \hat{\tau})^2}}\right) + \frac{b\hat{I}_{\text{max}}}{(1-d)\hat{\tau}} < 1,
\end{align}
where
\begin{align}
    \hat{T}^{\text{min}} &= \hat{T}^{\text{min}}_{\text{on}} + 1, \quad
    b = e^{-\frac{\hat{T}^{\text{min}}}{\hat{\tau}}}, \quad d = e^{-\frac{\hat{T}^{\text{min}}_{\text{on}}}{\hat{\tau}}}, \quad \hat{\tau} = \frac{\tau}{T_{\text{on}}}, \nonumber \\ 
    \hat{A}_{ub} &= \frac{A_{ub}}{m_1T_{\text{on}}}, \quad
    \hat{I}_{\text{max}} = \frac{I_{\text{max}}}{m_1T_{\text{on}}},\quad 
    \hat{\omega}_l = \frac{\omega_l T_{\text{on}}}{2\pi}.
\end{align}
\end{theorem}
The proof can be found in \cite{Avestruz2022b}.
$\hat{\tau}$ can be interpreted as the \emph{normalized time constant}.
$\hat{T}^{\text{min}}$, $\hat{T}^{\text{min}}_{\text{on}}$, 
$\hat{A}_{ub}$, $\hat{I}_{\text{max}}$, and $\hat{\omega}_l$ can be interpreted as the normalized minimum switching period, minimum on time, interference amplitude, maximum inductor current, and lower bound on the interference, respectively.

Theorem \ref{theorem:taomin4stability} provides a sufficient condition for stability. If $\tau$ satisfies this condition, global asymptotic stability is guaranteed.
% We show that our sufficient condition is very close to the sufficient and necessary condition. This means that the range of $\tau$ calculated from our sufficient condition is not too conservative.
This bound on $\tau$ depends on the maximum inductor current $I_{\text{max}}$, minimum on time $T^{\text{min}}_{\text{on}}$, and the interference amplitude and frequency bounds.
% Given the appropriately worst-case choice of $I_p$ and $I_c$, the sufficient bound is sufficient and necessary.
%It is an affine function of the amplitude of interference.
% close?? sufficient and necessary--tight bound; claim is that the sufficient is tight also?
\begin{theorem} \label{theorem:taomin4stability}
A current control loop using constant off\nobreakdash-time has a minimum on time $T_{\text{on}}^{\text{min}}$ and fixed off time $T_{\text{off}}$. The time constant of the first\nobreakdash-order, low\nobreakdash-pass filter is $\tau$. The interference $w(t)$ satisfies Definition\,2 in the Part I article \cite{cmpartone2022}.
The bounds on $\tau$ to guarantee the global asymptotic stability of the current control loop is
\begin{align} %\label{thoerem:filterstab}
 k_0\frac{1}{\hat{\tau}} + k_1 \frac{\hat{A}_{ub}}{\hat{\tau}}+ k_2 \frac{\hat{A}_{ub}}{\hat{\tau}\sqrt{1 + (2\pi\hat{\omega}_l\hat{\tau})^2} } < \frac{1}{2},
\end{align}
and
\begin{align} %\label{thoerem:filterstab}
 k_3 \frac{\hat{I}_{\text{max}}}{\hat{\tau}} + \frac{\hat{A}_{ub}}{\hat{\tau}} + \frac{\hat{A}_{ub}}{\hat{\tau}\sqrt{1 + (2\pi\hat{\omega}_l\hat{\tau})^2}} < \frac{1}{2},
\end{align}
where
\begin{align}
    k_0 & = \frac{d(\hat{T}_{\text{on}}^{\text{min}} + \hat{\tau} d -\hat{\tau})}{(1-d)^2}, \quad k_1 = \frac{1}{(1-d)}, \; \hat{T}^{\text{min}} = \hat{T}^{\text{min}}_{\text{on}} + 1, \nonumber \\
    b & = e^{-\frac{\hat{T}^{\text{min}}}{\hat{\tau}}}, \quad d = e^{-\frac{\hat{T}^{\text{min}}_{\text{on}}}{\hat{\tau}}}, \quad \hat{\tau} = \frac{\tau}{T_{\text{on}}}, \quad \hat{A}_{ub} = \frac{A_{ub}}{m_1T_{\text{on}}}, \nonumber \\
    k_2 & = 1 + \frac{(1+d)d}{(1-d)^2}, \quad k_3 = \frac{d-b}{(1-d)^2}, \quad \hat{I}_{\text{max}} = \frac{I_{\text{max}}}{m_1T_{\text{on}}}, \nonumber \\
    \hat{\omega}_l & = \frac{\omega_l T_{\text{on}}}{2\pi}.
\end{align}
\end{theorem}
The proof can be found in \cite{Avestruz2022b}.
\begin{comment}
do not have continuous deformation in psi!  why is the case?  because you cannot find attributes for the interference that enforce continuous deformation in psi.

interference: 
1) continuous phase of something periodic -- highly narrow class

T and psi are smooth nonlinearities.

Find worst-case nonlinearity in QS mapping from worst-case frequency and worst-case amplitude.

Does the worst case QS nonlinearity correspond to the worst-case slope in psi?

\red{if the design for control performance is based on small signal, then for all ic has to be small signal stable. you need to restrict psi and/or G(z) so the small signal stability for all ic is enforced.  wide-sense small signal stable and large stable at the same time.
(1) choose G(z) so it is stable for all psi'
(2) step 1: G(z) is not conditionally stable over psi'
(3) step 2: can check that it is stable for max psi'}
\end{comment}
Although these guarantees are large-signal, we accrue intuition on the stability through linearization. 
At the operating point determined by the peak inductor current command $I_c$, the actual peak inductor current $I_p$, the actual valley inductor current $I_v$, at on time $T_{\text{on}}$, we linearize the system (\ref{eqn:filter_nonlin_sys}) and (\ref{eqn:filter_nonlin_sys_fb}) as
\begin{align} \label{eqn:filter_lin_sys2}
\tilde i_p[n] = \; & \tilde i_p[n-1]  +m_1 \, \tilde t_{\text{on}}[n], \nonumber \\
\tilde i_c [n] = \; & q(T_{\text{on}}) q(T_{\text{off}}) \,\tilde i_c [n-1] +  c_1 \, \tilde i_p [n] +  c_2 \, \tilde t_{\text{on}}[n],
\end{align}
where 
\begin{align} \label{eqn:prarameterszdomain}
    c_1 = &  u(t) {\ast} h(t)\bigg\rvert_{t = T_{\text{on}}}, \nonumber \\
    c_2 = & -\frac{d\,q\,(t)}{d\,t}\bigg\rvert_{t = T_{\text{on}}} q(T_{\text{off}})I_c + h(T_{\text{on}})I_v \nonumber \\
    & + \frac{d\,w(t)u(t){\ast}h(t)}{d\,t} \bigg\rvert_{t = T_{\text{on}}}.
\end{align}
System (\ref{eqn:filter_lin_sys2}) is represented in the block diagram in Fig.\;\ref{fig:filteredccb_fb},
% \begin{figure}
%     \centering
%     \includegraphics[width = 8cm]{Figure/section3/filter/filter_fb.pdf}
%     \caption{Block diagram representation of the current control loop of PCC with filter}
%     \label{fig:filteredccb_fb}
% \end{figure}
with the plant
\begin{align} \label{eqn:gzofcotcm_part2_filter}
G(z) = \frac{1-z^{-1}}{m_1}.
\end{align}
Compared to the standard Lure representation in Fig.\,14(b) of the Part I article \cite{cmpartone2022}, there is an additional gain block 
\begin{align}
K = \frac{1}{1-e^{-\frac{T_{\text{on}}}{\tau}}}.
\end{align}
There also an additional pole-zero pair
\begin{align}
    F(z) = 1 - e^{-\frac{T}{\tau}}\,z^{-1}.
\end{align}
The effect of interference is shown in the feedback path $\psi_1$ in Fig.\,\ref{fig:filter_cotcm}
\begin{align}
    %  \psi_1 = \frac{w(T_{\text{on}})}{\tau} - \frac{e^{-\frac{T_{\text{on}}}{\tau}}}{\tau}\int_{-\infty}^{+\infty} \frac{W(\omega)}{j\omega}\,d\omega.
    \psi_1 & = \int_{-\infty}^{+\infty} \frac{j\omega W(\omega) e^{j\omega T_{\text{on}}}}{1 + j \omega \tau}\,d\omega - \frac{e^{-\frac{T_{\text{on}}}{\tau}}}{\tau}\int_{-\infty}^{+\infty} \frac{W(\omega)}{1+j\omega \tau}\,d\omega.
\end{align}
Negative $\psi_1$ can result in positive feedback and might destabilize the current control loop.
As the filter time constant $\tau$ is increased, the filter better attenuates the interference. 
We observe there is an another feedback path $\psi_2$ in Fig.\,\ref{fig:filter_cotcm}, which is derived from the static mapping
\begin{align}
     \psi_2 = - \frac{e^{-\frac{T}{\tau}}}{\tau}I_c + \frac{e^{-\frac{T_{\text{on}}}{\tau}}}{\tau}I_v.
\end{align}
This feedback path is a function of the actual current $I_p$ and inductor current ripple $m_2T_{\text{off}}$. 
% {\color{red}
% earlier section we said that the slope was separable in the static mapping because it did not affect the stability of the current loop. ---> only true for the ideal static mapping
% control conditioning -- now slope matters-->  matters for non-ideal static mapping.
% }
A large $I_p$ or small $m_2T_{\text{off}}$ can result in positive $\psi_2$.
Positive $\psi_2$ means a negative feedback path, which can partially cancel the positive feedback path of $\psi_1$. It is worth noting that $\psi_2$ is the result of the trapezoidal shape of the current sensor waveform.

We visualize the effect of the filter on the stability of the current control loop through the root locus.
$P(z)$ in Fig.\;\ref{fig:filteredccb_fb} is outside of the loop, hence is not included in the root locus. 
% Moreover, in the worst case, the valley current $I_p-m_2T_{\text{off}}$ is so small that the effect of $\psi_2$ is much smaller than $\psi_1$. 
We collapse $\psi_1$ and $\psi_2$ in the root locus as a single gain.
The root locus of the current control loop is shown in Fig.~\ref{fig:filter_cotcm}. 
We observe that the low\nobreakdash-pass filter for positive feedback in Figs.~\ref{fig:filter_cotcm} (a) and (b) move the location of the worst\nobreakdash-case closed\nobreakdash-loop pole further to the right, hence improving stability margin.  Likewise, for negative feedback in Figs.~\ref{fig:filter_cotcm} (c) and (d), the closed\nobreakdash-loop pole moves further to the left. 
In this way, the filter guarantees the closed\nobreakdash-loop poles to always stay inside the unit disk. Therefore stability is guaranteed.

We next show the quantitative relationship between the interference and the location of the closed\nobreakdash-loop pole. From (\ref{eqn:filter_lin_sys2}), the closed\nobreakdash-loop transfer function is 
% \begin{align} \label{iltf1}
% C_2(z) = \frac{\beta (1-b^{i}_1  z^{-1})}{1-a^{i}_1  z^{-1}}
% \end{align}
% \begin{align}
%     a_1^i &= 1 - \frac{m_1(1-e^{-T_{\text{on}}/\tau})}{m_1(1-e^{-T_{\text{on}}/\tau}) + (I_p - m_2T_{\text{off}} - I_c e^{-T_{\text{off}}/\tau}) \frac{e^{-T_{\text{on}}/\tau}}{\tau} + \frac{A\, \text{cos}(\omega T_{\text{on}}+\varphi)}{\tau}}, \nonumber \\
%     b^{i}_1 & =  e^{-T/\tau},  \nonumber \\
%     \beta & = \frac{m_1}{m_1(1-e^{-T_{\text{on}}/\tau}) + \left(I_p - m_2T_{\text{off}} - I_c  e^{-T_{\text{off}}/\tau}\right) \frac{e^{-T_{\text{on}}/\tau}}{\tau} + \frac{A\, \text{cos}(\omega T_{\text{on}}+\varphi)}{\tau}}.
% \end{align}
\begin{align}
     \label{eqn:iltf1} 
     C_2(z) = \frac{\beta (1-b  z^{-1})}{1-a  z^{-1}},
\end{align}
where
\begin{align} %\label{eqn:filter_ztransform}
    a & = 1 - \frac{m_1}{m_1 +  \frac{\psi_1 + \psi_2}{1-d}},
    % a & = 1 - \beta, \quad \beta = \frac{m_1}{s}, 
    \quad b = e^{-\frac{T}{\tau}}, \quad d = e^{-\frac{T_{\text{on}}}{\tau}}, \nonumber \\
    % s & = m_1 +  \frac{\psi_1 + \psi_2}{1-d}, \nonumber \\
    \psi_1 & = \int_{-\infty}^{+\infty} \frac{j\omega W(\omega) e^{j\omega T_{\text{on}}}}{1 + j \omega \tau}\,d\omega - \frac{e^{-\frac{T_{\text{on}}}{\tau}}}{\tau}\int_{-\infty}^{+\infty} \frac{W(\omega)}{1+j\omega \tau}\,d\omega,
     \nonumber \\
     \psi_2 & = - \frac{e^{-\frac{T}{\tau}}}{\tau}I_c + \frac{e^{-\frac{T_{\text{on}}}{\tau}}}{\tau}I_v.
\end{align}
% From (\ref{eqn:filter_nonlin_sys}) to (\ref{eqn:filter_lin_sys2}), we use the following algebraic deformation,
% \begin{align} 
%     \label{eqn:uhconv} u(t){\ast}h(t) &= \left(1 -e^{-\frac{t}{\tau}}\right)u(t) = (1 - d)u(t), \\
%     \label{eqn:sinapprox} A\,\omega\, \text{cos}(\omega  t+\varphi) u(t) {\ast} h(t) + A\,\text{sin}\varphi h(t) &=  \frac{A \,\omega \,\text{sin}(\omega t + \varphi + \delta)}{\sqrt{1+\omega^2\tau^2}} - \frac{A\, \omega \,\text{sin}( \varphi + \delta)}{\sqrt{1+\omega^2\tau^2}} e^{-\frac{t}{\tau}} + \frac{A\, \text{sin} \varphi}{\tau} e^{-\frac{t}{\tau}} \nonumber \\
%      & \approx \frac{A\, \text{sin}(\omega t + \varphi + \delta)}{\tau} - \frac{A\, \text{cos}( \varphi + \frac{\delta}{2})\, \text{sin} (\frac{\delta}{2})}{\tau}\, e^{-\frac{t}{\tau}}  \nonumber \\
%      &\approx \frac{A\,\text{sin}(\omega t + \varphi)}{\tau},
% \end{align}
% where $\delta = \text{tan}^{-1}(1/\omega \tau)$. Equation (\ref{eqn:sinapprox}) utilize the assumption that $\omega \tau >> 2 \pi$ and $\delta \rightarrow 0$.
% \begin{align}
%     s_1 &= m_1 (1-e^{-T_{\text{on}}/\tau}) + \frac{A sin(2\pi f t+\phi(f))}{\tau +1/2 \pi f} + \frac{(I_p - m_2T_{\text{off}}-I_c e^{-T_{\text{off}}/\tau})}{\tau}e^{-T_{\text{on}}/\tau} \\
%     \gamma & = 1 - e^{-\frac{T_{\text{on}}}{\tau}} \quad
%     a  = \frac{(I_p - m_2T_{\text{off}}-I_c e^{-T_{\text{off}}/\tau})}{\tau}e^{-T_{\text{on}}/\tau} \quad  b^{i}_1  =  e^{-T/\tau}
% % \end{align}
% Therefore, in $z$-domain, as shown in Fig.  \ref{fig:filter2}, the filter shrinks the poles which were outside of the unit disk into the unit disk. 

We use the settling and overshoot metrics in (\ref{eqn:settlecycle1_part2}) and (\ref{eqn:overshoot1_part2}) respectively from \cite{cmpartone2022}, to quantitatively illustrate the relationship between these metrics and the pole locations.
The settling cycles
\begin{align} \label{eqn:settlecycle1_part2}
    N_w = \text{max} \bigg\{\bigg|\frac{4}{\text{ln}(|a_{\text{min}}|)}\bigg|,\bigg|\frac{4}{\text{ln}(|a_{\text{max}}|)}\bigg|\bigg\}.
\end{align}
The \emph{worst-case overshoot}
\begin{align} \label{eqn:overshoot1_part2}
    O_w = \text{max}\{-a_{\text{min}},0\}, 
\end{align}
where $O_w$ is expressed in percentage form in this paper.

\begin{figure}[htp]
\centering
\subfigure[The worst\nobreakdash-case settling $N_w$ decreases first and then increases with the time constant $\hat{\tau}$ of the filter. The interference frequency $\hat{\omega} = 2$.]{
\includegraphics[width=7.5 cm]{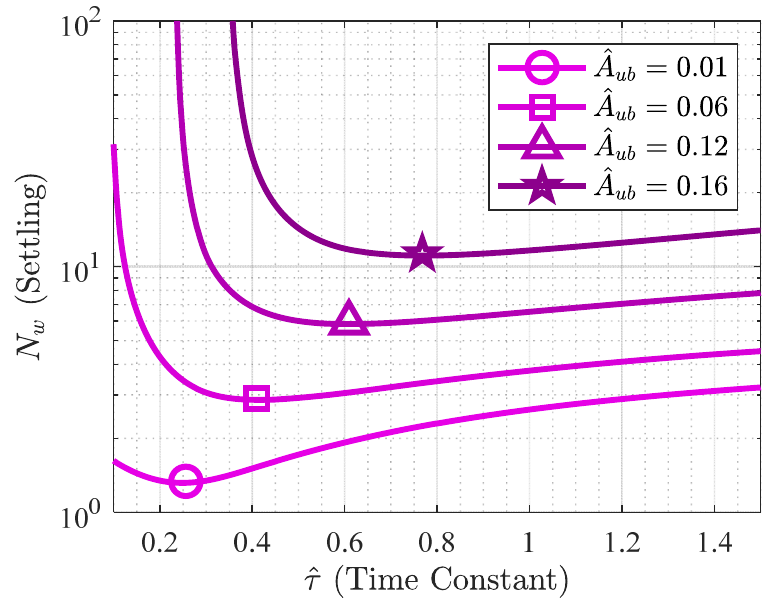} \label{fig:filter_settlingtime}
}
\subfigure[The worst\nobreakdash-case overshoot $O_w$ decreases first and then increases with the time constant $\hat{\tau}$ of the filter. The interference frequency $\hat{\omega}= 2$.]{
    \includegraphics[width=8 cm]{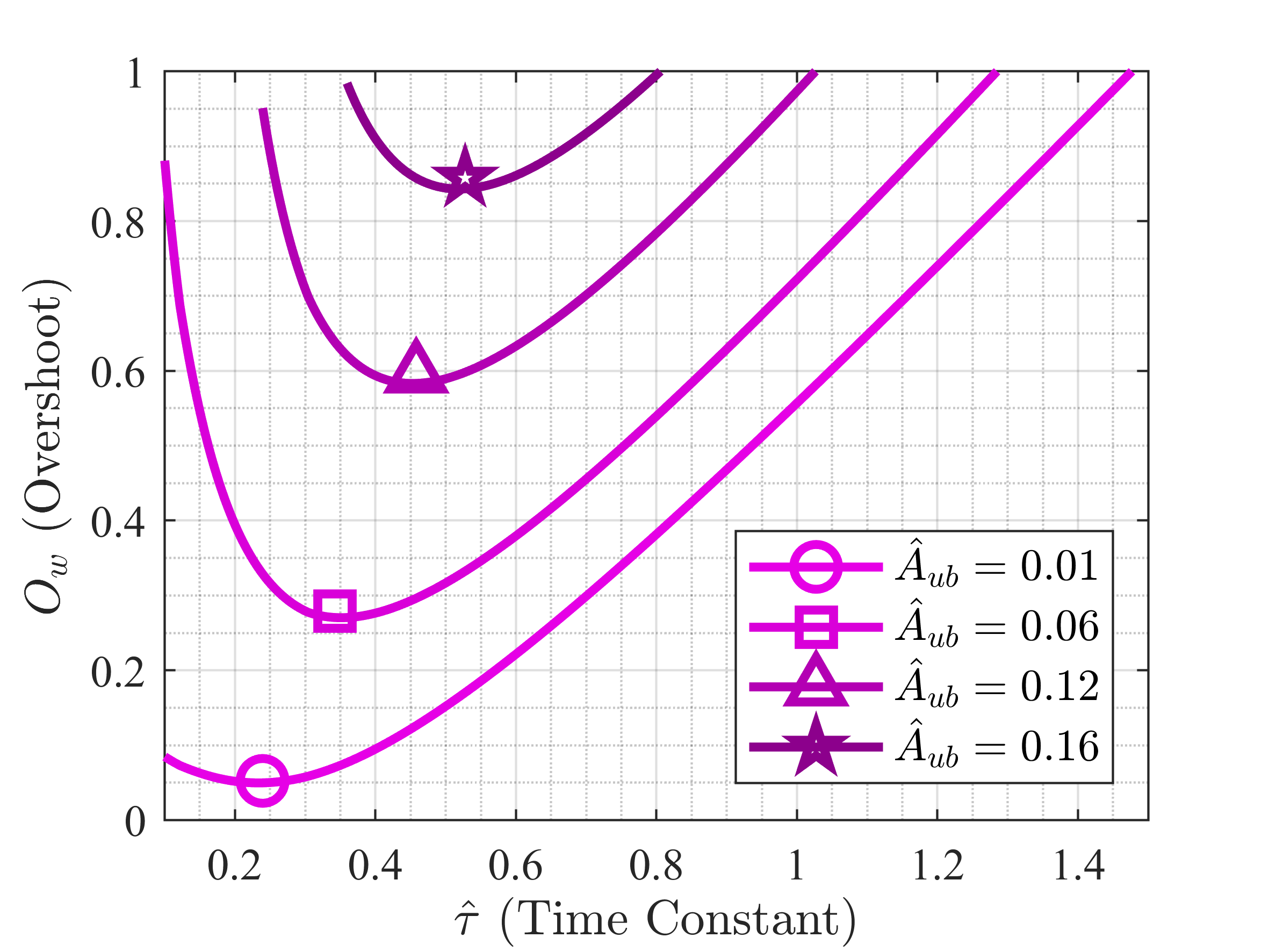} 
    \label{fig:filter_overshoot}
}
\caption{The design diagram for filter given different interference levels. The theoretical curves which are represented by the dotted line matches the simulated curves, which are represented by the solid line.}
%\label{fig:subfigureExample}
\end{figure}

For different operating points, the pole $a$ is within the range \mbox{$[a_{\text{min}},a_{\text{max}}]$}.
We observe that a small $\tau$ does not provide enough attenuation on the interference whereas a too\nobreakdash-high $\tau$ distorts the original current ramp. 
Another observation is that a big $\tau$ slows down the zero in (\ref{eqn:iltf1}). A slow zero causes a long\nobreakdash-tail settling and overshoot problem in the transient response. 
%So there exists a feasible interval $[\tau_{\text{min}}, \tau_{\text{max}}]$ for selecting $\tau$.
% such that for all $\tau \in [\tau_{\text{min}}, \tau_{\text{max}}]$, all the poles of the current control loop can stay inside the unit disk.
Figure\,\ref{fig:filter_settlingtime} and Fig.\,\ref{fig:filter_overshoot} show how the settling and overshoot change with the time constant. We fix the interference frequency at twice the switching frequency \footnote{For the constant on(off)\nobreakdash-time operation, this means the reciprocal of on(off) time, i.e. the fastest switching frequency.} and vary the interference amplitude.
The theoretical relationships are represented by the dotted line and the simulated relationships are represented by the solid line. 
% The assumption (\ref{eqn:timescaleassumption}) is plotted as the dot-dash line. The half space on the right side of the dot-dash line satisfies the assumption.
We observe that the theoretical curves perfectly match the simulation curves. 
% It is worthwhile to mention that even on the left side of the dot-dash line where the assumption (\ref{eqn:timescaleassumption}) breaks, the theoretical curves can still provide a decent estimation on the settling and overshoot.

The settling first decreases with the time constant $\hat{\tau}$ because the filter attenuates the interference. However, the filter also distorts the original current ramp. With the increase of $\hat{\tau}$, this distortion effect becomes more severe, hence the settling starts to increase with $\hat{\tau}$. Similarly, the overshoot decreases first with $\hat{\tau}$ because the interference is suppressed. However, the filter also introduces a zero in the transfer function (\ref{eqn:iltf1}). This zero becomes slower and causes large overshoot as $\hat{\tau}$ increases. 

In conclusion, the important take\nobreakdash-away messages for designers are: (i) there exists the optimal $\hat{\tau}$ to minimize overshoot and settling; (ii) the optimal $\hat{\tau}$ increases as the amplitude of interference increases, which is supported by our design diagram. 
% 3) the optimal $\hat{\tau}$ does not change as the frequency of interference changes, which is supported by our theory. 
% The step\nobreakdash-by\nobreakdash-step \red{design guide} for filter is shown in Section \ref{subsection:lpf}.

% \subsubsection{Analysis on Constant On-time current control loop}
To extend the analysis to the constant on-time current control loop, $G(z)$ is transformed to
\begin{align}
    G(z) = \frac{1-z^{-1}}{m_2},
\end{align}
and $\psi_2$ becomes
 \begin{align}
     \psi_2 = -\frac{d}{\tau}I_p - \frac{b}{\tau}I_c.
 \end{align}
Note that $\psi_2<0$, hence the feedback path $\psi_2$ is a pure positive feedback and it always shrinks the stability margin. Therefore, the filter helps more on the constant off\nobreakdash-time control than the constant on\nobreakdash-time control. For designers who use the filter in constant on\nobreakdash-time control, the author suggests that it is better to choose the time constant of the filter to be faster than the switching period to avoid the harmful effect of the  $\psi_2$ feedback path. The simplified root locus of the current control loop is identical to Fig.\;\ref{fig:filter_cotcm}.

\begin{figure}
    \centering
    \includegraphics[width = 8cm]{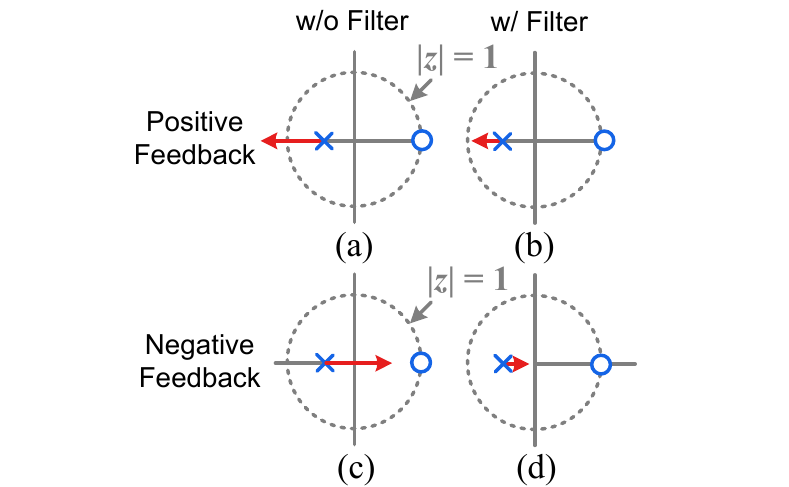}
    \caption{Small-signal root locus of the fixed-frequency current control loop with the low-pass filter. The sign of the feedback is determined by the interference. In positive feedback, low-pass filter decreases the loop gain to increase the stability margin. In negative feedback, the closed\nobreakdash-loop pole is moved further to the left by filter.}
    \label{fig:fpcc_filter}
\end{figure}

% \subsubsection{Analysis on Fixed-frequency current control loop}
The analysis method can be extended to fixed-frequency peak current-mode control with a $G(z)$ given by
\begin{align}
    G(z) &= \frac{1-z^{-1}}{m_1 + m_2 z^{-1}}, \\
    \psi(x) &= w(x + DT) - w(DT).
\end{align} 
and $\psi_2$ in the feedback loop given by
\begin{align}
     \psi_2 = \frac{d}{\tau}I_v - \frac{b}{\tau}I_c(1 - z^{-1}).
\end{align}
%We take the fixed-frequency peak current-mode control as an example. 
% We substitute the (\ref{eqn:filter_nonlin_sys}) by (\ref{eqn:ffpccip}) and substitute (\ref{eqn:filter_nonlin_sys_fb}) by 
% \begin{align}
%   i_c [n] &= 
%   e^{-\frac{t_{\text{on}}[n]}{\tau}}
%   e^{-\frac{T_{\text{off}}[n-1]}{\tau}} i_c[n-1] + \left(i_v[n] + m_1 t + A \, \text{sin}(\omega  t+\varphi) \right) u(t) {\ast} h(t)\bigg\rvert_{t=t_{\text{on}}[n]}.
% \end{align}
We observe that in constant on(off)\nobreakdash-time control, $\psi_2$ is a pure gain; however, in fixed\nobreakdash-frequency control, $\psi_2$ introduces a pole at $z=0$ and a zero. The simplified root locus of the current control loop is shown in Fig.\;\ref{fig:fpcc_filter}. 
We have shown a rigorous analytical way to design a filter for control conditioning. In the \mbox{traditional} way of designing for signal conditioning, low-pass filters are commonly chosen to cut off after the switching frequency, but far before the higher frequency interference band.
This makes the filters unable to effectively suppress interference whose spectrum is near or below the switching frequency.
% In designing filters for control conditioning, the filter time constant is selected through an accurate design algorithm in Section\,\cite{subsection:lpf}.
The cut\nobreakdash-off frequency of the filter can be well\nobreakdash-below the switching frequency yet still result in stability and good transient performance.
Even when the bandwidth of the filter is particularly low that the ramp is significantly distorted, the stability of the current control loop can still be guaranteed. 
%Low-pass filter has a symmetric effect on the peak current\nobreakdash-mode control and valley current\nobreakdash-mode control. A low-pass filter with low cut-off frequency compared to switching frequency has a severe stabilizing effect on the valley current\nobreakdash-mode control.

\subsection{Comparator-Overdrive-Delay Conditioning} \label{section:opd}

In this section, we introduce a new idea using the comparator overdrive delay in real implementations of comparators to condition for interference; it is a dual-use of the comparator.  Comparator overdrive delay is a propagation delay that is dependent on the input voltage difference. In an integrated circuit, comparator overdrive delay can be made to be real-time programmable. 

\begin{figure}
    \centering
    \includegraphics[width = 6cm]{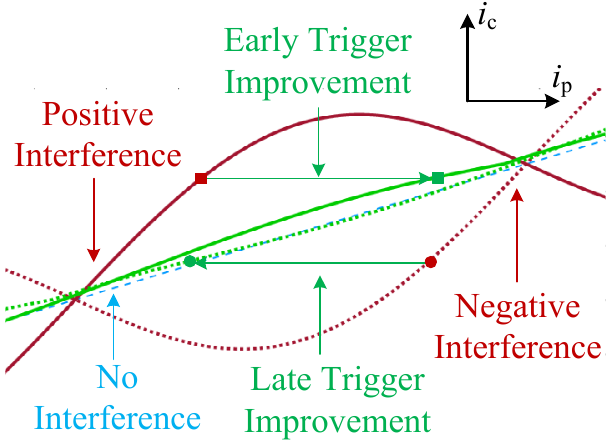}
    \caption{Comparator overdrive decreases the degree of nonlinearity in the static current mapping.}
    \label{fig:comp_w_inf}
\end{figure}

\begin{figure}
    \centering
    \includegraphics{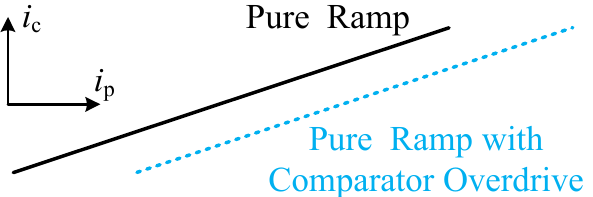}
    \caption{Comparator overdrive does not affect the nonlinearity of the static current mapping and only causes an offset error.}
    \label{fig:comp_no_inf}
\end{figure}

Like other control conditioning methods, using the comparator overdrive delay can repair both the discontinuity and nonlinearity problems in the static mapping along with improving the transient performance of dynamical mapping.  Comparator overdrive decreases the degree of nonlinearity in the defective static mapping shown in Fig.\,\ref{fig:comp_w_inf}.  For an ideal current sensor output ramp, using comparator overdrive delay does not create a nonlinearity, but instead introduces an offset in the static mapping, as shown in Fig.\,\ref{fig:comp_no_inf}.  This offset error does not affect the stability and transient performance of the current control loop.  Rather, it can be separated so that it is outside of the current control loop and thus becomes a disturbance for the outer voltage control loop, which is often addressed by an integrator in the controller.
In practice, the lower bound on the interference frequency $\omega_{l}$ can be chosen as within the bandwidth of the outer loop.
% As long as the delay caused by the comparator overdrive is not longer than the switching period, the control bandwidth of voltage-control loop will not be affected.

For an ideal current sensor output ramp, comparator \mbox{overdrive} delay conditioning introduces a fixed delay in the dynamical mapping. For fixed\nobreakdash-frequency power converters, this overdrive delay, together with other significant delays, which are caused by the blanking, subthreshold and signal propagation, should not exceed the duty ratio limits. Variable frequency power converters do not have this limitation.
%In high-frequency dc-dc converters, it is not accurate enough to just model the comparator as an ideal logic comparator.
% A new comparator model considering the overdrive propagation delay

In real voltage comparator implementations, the output of the comparator does not change instantaneously when the input difference crosses the voltage threshold, hence causing a delay.  The \emph{input overdrive} is defined as this voltage difference after the threshold is crossed, but before the output changes state.  The input overdrive can be positive or negative, \mbox{depending} on whether the signal is positive\nobreakdash-going or negative\nobreakdash-going.  For example, for peak current detection, input overdrive is when the current sensor voltage is above the voltage threshold set by the current command; for valley current detection, it is when the sensor voltage is below. 

The delay time depends on the input overdrive in what is known as \emph{comparator overdrive delay}, which is typically shown in datasheets.\footnote{The comparator overdrive delays that are indicated in datasheets often refer to the delays from step changes in the input.  Other types of inputs can be inferred from the datasheet delays.}  In the rest of this section, we present comparator overdrive conditioning, without loss of generality, in the context of current control loops that use peak current sensing, which are positive\nobreakdash-going input signals.
\begin{figure}
    \centering
    \includegraphics[width=8 cm]{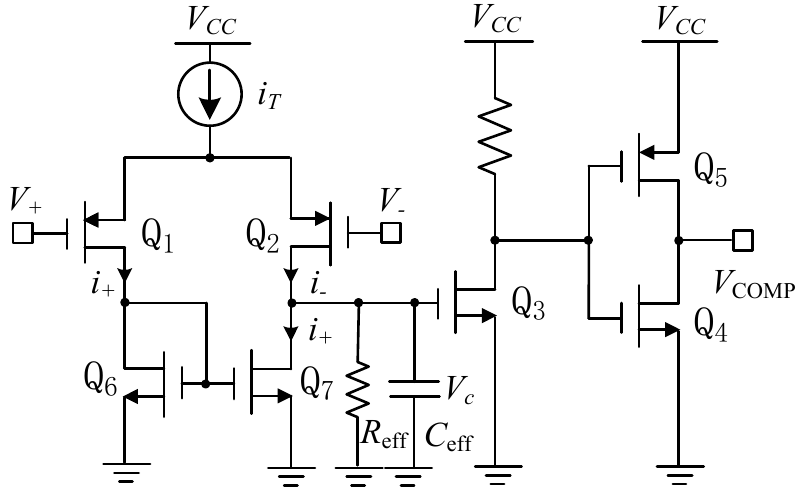}
    \caption{\label{fig:physicalmodel} A schematic of a simple three\nobreakdash-stage comparator.}
\end{figure}
% To take the variable delay into consideration, we show the following variable delay comparator model: 

%\subsubsection{Modeling the Comparator Overdrive}
%To model the comparator overdrive, 
To illustrate our model for comparator overdrive, we \mbox{examine} a simple three\nobreakdash-stage comparator shown in Fig.\;\ref{fig:physicalmodel}. It contains a differential\nobreakdash-pair front end (Stage I), a \mbox{common}\nobreakdash-source amplifier (Stage II), and a logic inverter output (Stage III).
\begin{figure}
    \centering
    \includegraphics[width=8.5cm]{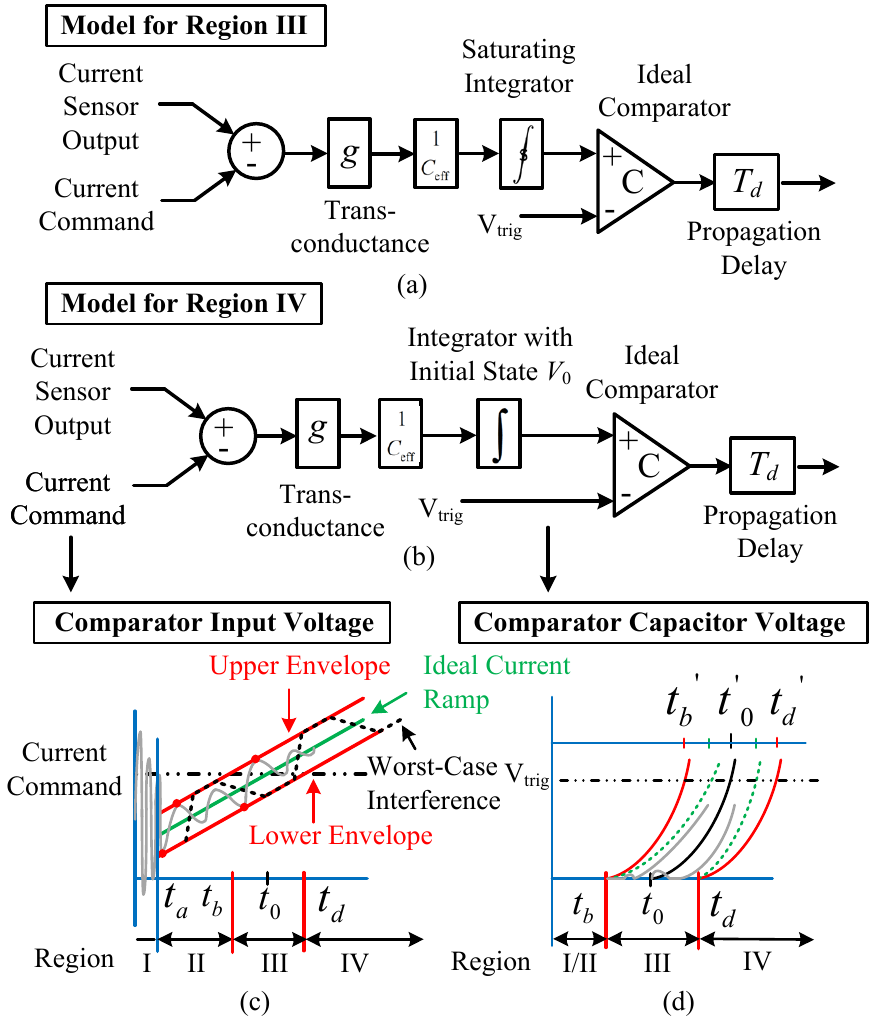}
    \caption{\label{fig:mathmodel}
(a) Comparator overdrive model for Region III.
(b) Comparator overdrive model for Region IV.
(c) Comparator input voltage. 
The four regions are divided by the time instants $t_a$, $t_b$, and $t_d$.
The blanking ends at $t_a$.
The upper envelope and lower envelope crosses the current command at $t_b$ and $t_d$, respectively.
(d) Comparator capacitor voltage.
The capacitor voltage trajectory of the upper envelope, ideal current ramp, and lower envelope cross the threshold $V_{\text{trig}}$ at $t_b^{'}$, $t_0^{'}$, and $t_d^{'}$, respectively.
The capacitor voltage trajectory given the worst\nobreakdash-case interference is bounded between the green dashed line.
Hence the comparator output can not trigger earlier than $t_b^{'}$ or later than $t_d^{'}$.}
\end{figure}

We denote the equivalent capacitance at the output of Stage~I by $C_{\text{eff}}$. The comparator output toggles only if the voltage difference $V_{+} - V_{-}$ lasts long enough so that $C_{\text{eff}}$ is charged or discharged to cross the voltage threshold of Stage~II.
The current at the output of Stage~I is $i_{-} - i_{+}$, which can also be expressed as $g(V_{-} - V_{+})$, where $g$ is a nonlinear transconductance \cite{razavi2002design}. 
% This current charges the parallel $C_{\text{eff}}$ and $R_{\text{out}}$, which is the equivalent resistor at the output the Stage I.
Because of $R_{\text{eff}}$, which is the effective resistance at the output of Stage~I, the charging current for $C_{\text{eff}}$ is always smaller than $i_{-} - i_{+}$.
$R_{\text{eff}}$ effectively decreases the transconductance.  For the analysis, we choose a constant $G$, which is the largest small\nobreakdash-signal transconductance in the range of $g$.  Choosing the largest transconductance results in the lowest comparator overdrive delay and hence a conservative design for stability.
% Because the tail current $i_T$ is constant, this requires including $i_T$ as the saturation capacitor current in the control conditioning model.
Because $g$ is determined by $i_T$, overdrive delay can be programmed by changing this tail current $i_T$, for example in an integrated circuit design.

We formulate a model class based on practical current sensor waveforms from power electronics, for example in Fig.\,\ref{fig:noisecurrent_part2}. 
This class elicits a model that is straightforward in guaranteeing global stability.

This model class is characterized by four regions, which can be observed in Fig.\,\ref{fig:mathmodel}(c).  
Region I is the ``blanking'' region, where large and very fast but quickly decaying switching transients dominate the current sensor output.  
It is worth noting that in the many instances that this region is blanked, the blanking occurs after the output of the comparator to defer the peak current event detection. 
Region II is the ``subthreshold'' region where the worst\nobreakdash-case current sensor waveform is below the current command.
Region III is the ``threshold'' region where the current sensor waveforms are neither unambiguously below nor above the current\nobreakdash-command threshold; in this region, the waveform can cross the threshold multiple times.
Region IV is the ``overdrive'' region where the current sensor waveform is always above the current command.
The qualifying restriction on Region II, III, and IV for this model class is that waveform is never below the minimum current command.  

In Regions I and II, we consider the comparator capacitor remaining in reset or equivalently, negatively saturated. In Regions III and IV, the capacitor integrates the current that is representative of the difference between the current command and the current sensor output; the mathematical formalization requires two different types of integrators to describe each region, as shown in Fig.\,\ref{fig:mathmodel}(a) and (b).  The integrator resets every switching cycle for the usual case that the overdrive delay of the comparator is smaller than the minimum off time.
% {\color{red} (Ask professor of this sentence is needed or not) need an explanation somewhere of Fig. 37d.}

The ideal current sensor output (without interference) is a ramp, which crosses the current command threshold at $t_0$.  After $t_0$, $C_{\text{eff}}$
%the capacitor on the comparator 
begins integrating; for the ideal current ramp, the capacitor voltage is a quadratic.  When the capacitor voltage crosses the trigger voltage $V_{\text{trig}}$ at $t_0$, which results from a combination of the gains and MOSFET thresholds (for example in Fig.\,\ref{fig:physicalmodel} for $\text{Q}_3$), the comparator output changes state at $t_0^{'}$.
The overdrive delay is from 
$t_0$ to $t_0^{'}$.
% the instant that the comparator input crosses the current command to the instant when the comparator output triggers.

The current sensor output with interference in our model class can be bounded from above by an \emph{upper envelope} and from below by a \emph{lower envelope} in Regions II and III.
The upper envelope crosses the current command earlier than the ideal current ramp by $\Delta t_b = t_0 - t_b$.
The lower envelope crosses the current command later than the ideal current ramp by $\Delta t_d = t_d - t_0$.
The upper and lower envelope  
have the same comparator overdrive delay as the ideal current ramp because these envelopes have the same slope as the ideal current ramp given the bounded amplitude assumption,
\begin{equation}
    t_0^{'} - t_0 \, = \, t_b^{'} - t_b \, = \, t_d^{'} - t_d.
\end{equation}
Although the overdrive delays are identical, the upper envelope triggers earlier than that of the ideal current ramp by \mbox{$\Delta t_b^{'} = t_0^{'} - t_b^{'}$} and the lower envelope triggers later by \mbox{$\Delta t_d^{'} = t_d^{'} - t_0^{'}$}.

The worst\nobreakdash-case interference waveform that can be contained by the upper and lower envelopes is a trapezoidal signal \cite{katznelson2004}.
The lower bound $\omega_l$ of the interference frequency is the fundamental of the trapezoidal signal. 
The largest slew rate of the trapezoidal signal is the upper bound of the Lipschitz constant ${\Lambda}_{ub}$.

Comparator overdrive conditioning decreases the trigger time deviation from the ideal ramp for waveforms within the envelope bounds. In Region III, the time deviation of the crossing event of the worst\nobreakdash-case interference from that of the ideal current ramp can range from $\Delta t^{'}_b$ earlier to $\Delta t^{'}_d$ later. In Region IV, the capacitor voltage trajectory given the worst\nobreakdash-case interference input is strictly bounded between the quadratic trajectories of the upper and lower envelope inputs.  This is a strict bound because the integral of the worst\nobreakdash-case trapezoid is bounded from above by $\pi A_{ub} / \omega_l$; hence, the comparator output cannot trigger earlier than $\Delta t_b^{'}$ nor later than $\Delta t_d^{'}$. The tradeoff of this control conditioning method is that longer overdrive delay becomes commensurate with smaller trigger time deviation.  This overdrive delay manifests as a limitation to an outer control loop, which might for example, control output voltage.
\begin{figure*}[ht]
\centering
\subfigure[Comparison of the model and experimental data on the overdrive delay of comparator LT1711. $\Delta v$ represents the input overdrive and $t_d$ represents the delay. The fitted trigger voltage --- time constant product $V_{\text{trig}}\tau = 6.102\,$ns$\cdot$mV, and $T_d = 4.198\,$ns.]
{\includegraphics[width = 0.3\textwidth ]{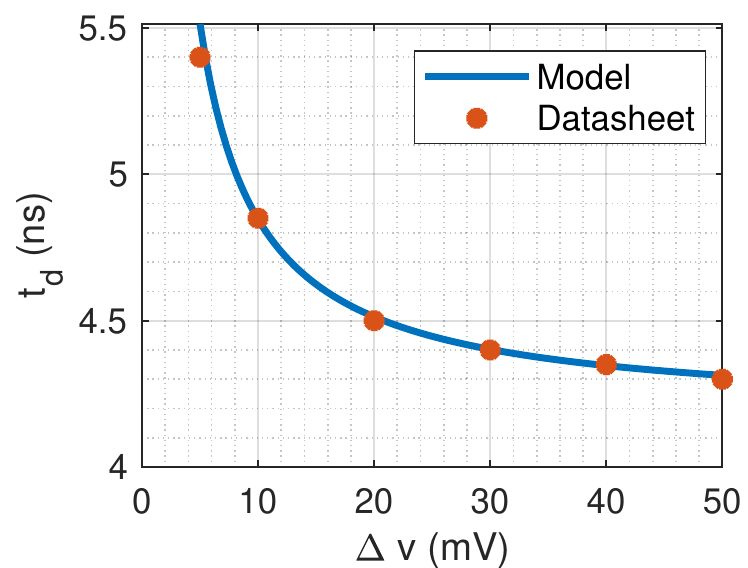}
\label{fig:simumodelcomp_lt1711}
}
\subfigure[Comparison of the model and experimental data on the overdrive delay of comparator AD8469. $\Delta v$ represents the input overdrive and $t_d$ represents the delay. The fitted trigger voltage --- time constant product $V_{\text{trig}}\tau = 113.3\,$ns$\cdot$mV, and $T_d = 24.75\,$ns.]
{\includegraphics[width = 0.3\textwidth ]{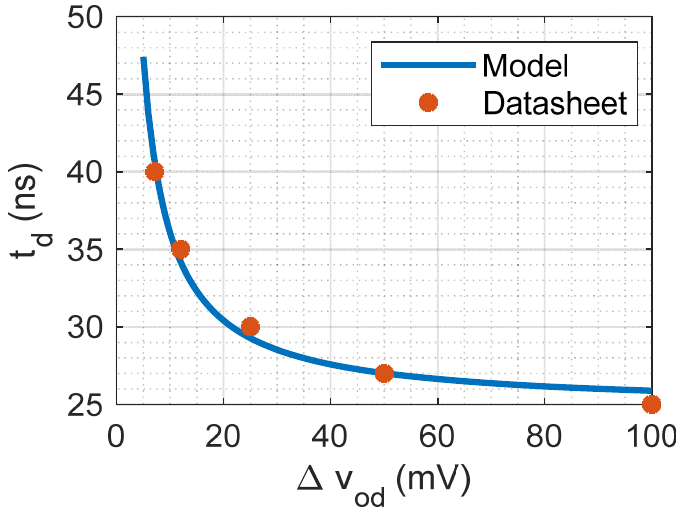}
\label{fig:simumodelcomp_ad8469}
}
\subfigure[Static mapping with different comparator overdrive delay. {\color{blue}-\,-\,-} is for the comparator with zero overdrive delay.  {\color{green}---} is for the comparator with medium overdrive delay. The jump discontinuity points are represented by {\color{red}$\cdots$}. 
  {\color{orange}-\,$\cdot$\,-} is for the comparator with long overdrive delay.]{ \includegraphics[width = 0.3\textwidth]{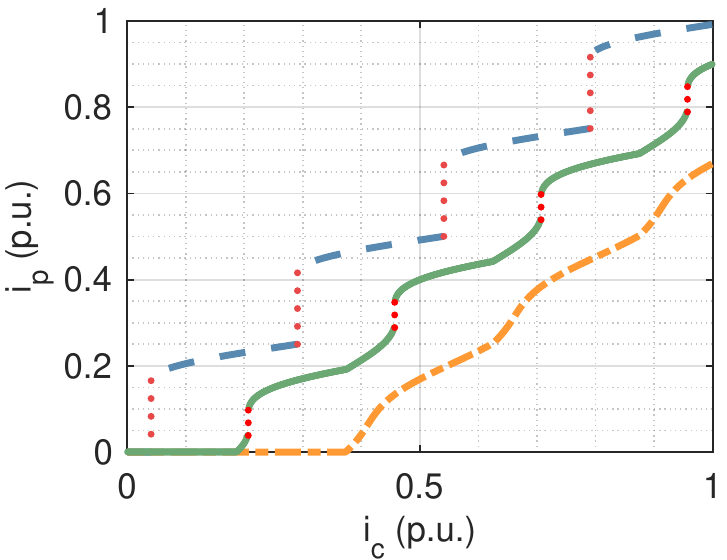} \label{fig:functionTwOPD}
}
\caption{Comparison of the model and experimental data on the overdrive delay.}
\end{figure*}
% \begin{figure}
%   \centering
%   \includegraphics[width=8 cm]{simumodelcomp.pdf}
%   \caption{\label{fig:odpopcomp} Comparison of the model and experimental data on the overdrive delay of comparator LT1711 \cite{Dual2001}. $\Delta v$ represents the input overdrive and $t_d$ represents the delay. The fitted trigger voltage --- time constant product $V_{\text{trig}}\tau = 6\,$ps$\cdot$V, and $T_d = 4.18\,$ns.
% %   Note that this not the only parameter set which can fit the experimental data.
%   \newline}
% \end{figure}
% \begin{figure}
%   \centering
%   \includegraphics[width=8cm]{functionTOPD.pdf}
%   \caption{\label{fig:functionTwOPD} Static mapping with different comparator overdrive delay. {\color{blue}-\,-\,-} is for the comparator with zero overdrive delay.  {\color{green}---} is for the comparator with medium overdrive delay.
%   The jump discontinuity points are represented by {\color{red}$\cdots$}. 
%   {\color{orange}-\,$\cdot$\,-} is for the comparator with long overdrive delay.}
% \end{figure}

%\subsubsection{Modeling the current controller with the %Comparator Overdrive}

%Region B/C boundary is worst-case delay.  Apply lower envelope to saturating integrator.

Delay and nonlinearity are the pertinent effects that determine stability and transient performance. Region I (blanking), Region II (subthreshold), and Region III (threshold) are modeled as pure delays in Fig.\;\ref{fig:mathmodel}(c). $T_d$ is a constant that encapsulates the circuit delays that are independent of the comparator input.  This model for comparator overdrive delay agrees well with the real\nobreakdash-world data \cite{Dual2001} shown in Fig.\;\ref{fig:simumodelcomp_lt1711} and Fig.\;\ref{fig:simumodelcomp_ad8469}.

The model for Region III (threshold) is shown in Fig.\;\ref{fig:mathmodel}(a).  The salient feature of this model is the saturating integrator,  
whose state and hence output is bounded both from above and below, but behaves like a linear integrator between these bounds.  An implementation of a comparator with this behavior is illustrated in Fig.\,\ref{fig:physicalmodel}, where $V_c$ is always above ground.  It is worth noting that the comparator circuit actually possesses this behavior in all four of the the model regions.  Stability guarantees can be proven by partitioning the comparator behavior into these regions and restricting the saturating integrator behavior to Region III.

This saturating integrator behavior is mathematically important in that the saturating integrator is a sublinear function, which enables the proof of continuity in the static mapping \cite{Avestruz2022a}.
% The saturating integrator can be found in Definition \ref{def:sat_int} in Appendix\,\ref{proof:gmaxcontinuity}.
\red{STOP} The behavior in Region III be represented by
    \begin{align} \label{eqn:IT_reg_3} V(t) =  \frac{1}{C_{\text{eff}}}\sint_{t_b}^{t_d} G\left(i_v[n] + m_1 t + w(t) - i_c [n] \right) dt.
\end{align}
We define $t_{fi}$ as the last instant in Region III when the output of the integrator $V(t)$ is zero.\footnote{We provide a short proof for this statement: the saturating integrator always has positive or zero state at any threshold crossing instants.
We start from the last instant and backwards check the integrator states at each threshold crossing instants.
We stop at the instant at which the integrator state is zero.
This algorithm is guaranteed to stop because the integrator state must be zero at the first threshold crossing instant.
Therefore, the stop instant is the $t_{fi}$ we want.}
\begin{align} \label{eqn:final_i_eqn_num_trick}
    t_{fi} \triangleq \left\{t^\prime ~ \rvert ~~ V(t^\prime) = 0 \,\text{and}\, V(t) > 0 ~~ \forall\,t > t^\prime\right\}.
\end{align}
% because ideal current ramp is monotonically increasing and the interference is bounded 
% For any instant when the integrator is zero, a negative error will keep the state still stay at zero and a positive error will immediately push the state to positive. Therefore, a threshold crossing.
% Because the current sensor output 

After $t_{fi}$, the behavior in Region III is a linear integrator with a voltage output $V_0$ at the end of Region III
\begin{align} 
    \label{eqn:v_0_rep} 
    % \frac{1}{C_{\text{eff}}} \sint_{t_{fi}}^{t_{d}} G\left(i_v[n] + m_1 t + w(t) - i_c [n] \right) dt \nonumber \\
      V_0 = \frac{1}{C_{\text{eff}}} \int_{t_{fi}}^{t_{d}} G\left(i_v[n] + m_1 t + w(t) - i_c [n] \right)\,dt.
\end{align}

The model for Region IV (overdrive), which represents a strictly positive integrator output is modeled in Fig.\;\ref{fig:mathmodel}(b).
The initial output $V_0$ of the integrator embeds the saturating integrator behavior in Region III.
$V_0$ is bounded from above by \mbox{$\frac{1}{2}m_1(t_b - t_d)^2$}, which is induced by the upper envelope of the current sensor output.
$V_0$ is bounded from below by the lower bound of the saturating integrator.

The comparator overdrive in Region IV can be represented by
    \begin{align} \label{eqn:IT_new} V_{0} + \frac{1}{C_{\text{eff}}}\int_{t_d}^{t_{\text{on}}[n]} G\left(i_v[n] + m_1 t + w(t) - i_c [n] \right) dt = V_{\text{trig}}.
\end{align}
% \begin{align}
%     \sint
% \end{align}
% The algorithm to find out the $G_{\text{max}}$, the maximum gain which make the equilibrium existence condition hold can be found in appendix.
In (\ref{eqn:IT_new}),
\mbox{$i_v[n] + m_1 t + w(t) - i_c [n]$} represents the error between the current sensor output and the current command.
% $\sint$ is a \emph{saturating integral operator} defined in Appendix \ref{proof:gmax4feasibility}.

% We assume that the maximum voltage difference between the comparator input $\delta v_{\text{comp}}$ is in 
% We denote the range of $g$  by $[G_{\text{min}},G]$.
% From our previous discussion of the Region III conditions, specifically,  $t_{fi}$ is the minimum solution of the equation $ m_1 t + f(t)  = i_c [n] - i_v[n]$ which satisfies
% We define the \emph{crossing time} $t_{fi}$ as the \emph{minimum solution} of the equation
% $ \label{eqn:DTC}
%  m_1 x + f(x)  = i_c [n] - i_v[n]$
% which satisfies
% \begin{align} \label{eqn:IT_nosat}
%     \frac{1}{C_{\text{eff}}} \int_{t_{fi}}^{t_{\text{on}}} g\left(i_v[n] + m_1 t + w(t) - i_c [n] \right) dt > 0, \quad \forall \; t_{\text{on}}>t_{fi}.  
% \end{align}
% By introducing $t_{fi}$, we convert the model in (\ref{eqn:IT}) which uses nonlinear saturating integral operator into the model in (\ref{eqn:IT_nosat}) which only uses linear integral operator.
By substituting (\ref{eqn:v_0_rep}) into (\ref{eqn:IT_new}), the comparator overdrive can be represented as
\begin{align} \label{eqn:IT} \frac{1}{C_{\text{eff}}}\int_{t_{fi}}^{t_{\text{on}}[n]} G\left(i_v[n] + m_1 t + w(t) - i_c [n] \right) dt = V_{\text{trig}}.
\end{align}
Together with the dynamics of the constant off\nobreakdash-time current control loop, the system using the comparator overdrive can be represented as
% Considering the actual peak current $i_p[n] = i_v[n] + t_{\text{on}}[n]$ and the peak current deviation $i_e[n] \triangleq i_c[n] - i_v[n]$ into (\ref{eqn:IT_nosat}), the current control loop using comparator overdrive can be represented as
% By substituting $i_p[n] = i_v[n] + m_1t_{\text{on}}[n]$ and $i_e[n] \triangleq i_c[n] - i_v[n]$ into (\ref{eqn:IT_nosat}), the current control loop can be represented as
\begin{subequations}
\label{eqn:opd_sys}
    \begin{align} 
    \label{eqn:ITlinear} \int_{t_{fi}}^{t_{\text{on}}[n]}
    & \left(i_v[n] + m_1 t + w(t) - i_c [n] \right) dt =  V_{\text{trig}}\tau, \\
    \label{eqn:forwardpath}
    & i_p[n] = i_p[n-1] - m_2T_{\text{off}} +m_1t_{\text{on}}[n],
    \end{align}
\end{subequations}
where the comparator time constant \mbox{$\tau = C_{\text{eff}}/G$} is the {\em design variable} for comparator-overdrive-delay conditioning. Equation (\ref{eqn:ITlinear}) describes the feedback path and (\ref{eqn:forwardpath}) describes the forward path. 

Proposition 1 in the Part I of the paper \cite{cmpartone2022} states that if the current sensor output is not monotonic, then the static current mapping is discontinuous.
A discontinuous static mapping can make the current control loop unstable.
We use $\tau$  as the metric for the comparator overdrive delay. \mbox{Theorem}\,\ref{theorem:gmaxcontinuity} below states that an appropriately designed comparator overdrive delay can make the static mapping continuous.
Figure\,\ref{fig:functionTwOPD} demonstrates Theorem\,\ref{theorem:gmaxcontinuity} by plotting three static mappings with different comparator overdrive delays. 
We observe that given small $\tau$, the static mapping is discontinuous.
As we increase $\tau$, the degree of discontinuity of the static mapping decreases. The static mapping becomes continuous if $\tau$ is large enough. Mathematically, the degree of discontinuity can be defined as the Lebesgue measure on the set of unreachable points.

% From Proposition\, \ref{th:equexcond}, 
% if the current sensor output is not monotonic, the static current mapping is not continuous anymore. 
% The current control loop may be unstable because of this discontinuity.
% The comparator overdrive delay decreases the degree of discontinuity of the static mapping so that it is continuous.
% We leverage the transconductance $G$ to adjust the overdrive delay. 
% Figure\;\ref{fig:functionTwOPD} illustrates the static current mapping $\mathcal{T}$ given the comparator with different transconductance and overdrive delay.
% % The Lebesgue measure of $U\!E$ exists and we denote it by $\lambda(U\!E)$.
% We note that as $G$ increases, the degree of the discontinuity of the static mapping decreases. 
% Therefore, we can expect that there exists a maximum error gain $G_{\text{max}}$ to guarantee the continuity condition and stability condition.

\begin{theorem} \label{theorem:gmaxcontinuity}
Given a constant off\nobreakdash-time current control loop with comparator overdrive delay, if the input is a ramp with slope $m_1$ and interference function $w(t)$, the condition to guarantee the continuous static current mapping is
\begin{align}
  V_{\text{trig}}\tau \ge m_1 K_3\!\left(\frac{|W(\omega)|}{m_1}\right).
\end{align}
\end{theorem}
The definition of the $K_3(\cdot)$ function as well as the proof of Theorem \ref{theorem:gmaxcontinuity} can be found in \cite{Avestruz2022a}.

Having satisfied the condition of continuous static mapping, we next show that the comparator overdrive delay allows the dynamical mapping to be stabilized.
Comparator overdrive conditioning attenuates the effect of interference by averaging, hence its performance improves with increasing interference frequency.  The averaging time interval is determined by $\tau$; smaller $\tau$ results in longer averaging time and hence a better interference attenuation.  Smaller $\tau$ contributes to a bigger delay, which means that interference attenuation trades off with transient performance. Theorem\,\ref{theorem:gmax4stability} describes the stability constraint on the design of $\tau$;
Theorem\,\ref{theorem:gmax4overdrivedelay} describes the constraint from the hardware limits on minimum on time.

Theorem\,\ref{theorem:gmax4stability} shows there exists a minimum comparator time constant, above which the current control loop is guaranteed to be globally asymptotically stable.
The stability of the dynamical mapping depends on the upper bound of the interference amplitude $A_{ub}$ and the lower bound of the interference frequency $\omega_l$.  Only a short overdrive delay is needed to stabilize the current control for small $A_{ub}$ and large $\omega_{l}$.  This is consistent with the behavior we described in Fig.\,\ref{fig:mathmodel}. 

% Given a continuous static mapping, the stability of the dynamical mapping is determined by the upper bound of the Lipschitz constant $L_{ub}$ of the interference.
% For the interference whose Lipschitz constant is outside of this bound, we show that an appropriately designed comparator overdrive delay can still guarantee the globally asymptotic stability of the current control loop. 
% The stability condition depends on the upper bound of the amplitude of the interference $A_{ub}$ and lower bound of the frequency spectrum of the interference $\omega_l$. 
% The smaller $A_{ub}$ is, smaller $\omega_l$ is or larger $G_{\text{max}}$ is, the better stability margin the current control loop has.
% This matches the qualitative analysis in Fig.\,\ref{fig:mathmodel}.
\begin{theorem} \label{theorem:gmax4stability}
Given a constant off\nobreakdash-time current control loop with comparator overdrive delay, if the input is a ramp with slope $m_1$ and interference function $w(t)$,
% maximum $T^{\text{max}}_{\text{on}}$ in the transient, 
the condition to guarantee a globally asymptotically stable dynamical mapping is
\begin{align}\label{eqn:vtrigtau_constraint}
  V_{\text{trig}} \tau \ge \frac{4A_{ub}^2}{m_1} + B,
\end{align}
where
\begin{align} 
  \tau = \frac{C_{\text{eff}}}{G}, \quad B = \int_{-\infty}^{+\infty} \Bigg| \frac{W(\omega)}{\omega}  \Bigg| \,d\omega \nonumber.
\end{align}
\end{theorem}
% The constraint (\ref{eqn:vtrigtau_constraint}) is derived from Theorem \ref{theorem:gloasystab}.
The proof of this Theorem is based on the observation that (\ref{eqn:ITlinear}) is an implicit nonlinear function from $t_{\text{on}}[n]$ to the deviation of peak current \mbox{$i_c[n] - i_p[n]$}. 
Therefore, the current\nobreakdash-control loop can be represented as a Lure system.
Equation (\ref{eqn:forwardpath}) describes the linear forward path and $G(z)$ can be found in (\ref{eqn:gzofcotcm_part2_filter}).
(\ref{eqn:ITlinear}) describes the nonlinear feedback path and $\psi(\tilde{t}_{\text{{on}}})$ is sector\nobreakdash-bounded.
The detailed proof can be found in \cite{Avestruz2022a}.
Equation (\ref{eqn:vtrigtau_constraint}) shows that a current control loop with large interference amplitude can be stabilized by a comparator with a large time constant $\tau$.

However, having a time constant $\tau$ that is too large results in an overdrive delay that may be too long. Long overdrive delay increases the minimum on time $T_{\text{on}}^{\text{min}}$ of the current control loop, hence slowing down transient response. Theorem~\ref{theorem:gmax4overdrivedelay} shows a rigorous upper bound on the overdrive delay.
% design of $\tau$ given the constraint on $T_{\text{on}}^{\text{min}}$.
\begin{theorem}
\label{theorem:gmax4overdrivedelay}
Given a constant off\nobreakdash-time current control loop with comparator overdrive delay, if the input is a ramp with slope $m_1$ and interference function $w(t)$,
the maximum comparator overdrive delay $t_{od}^{\text{max}}$ is
\begin{align}\label{eqn:tau_constraint}
    t_{od}^{\text{max}} = \frac{A_{ub}}{m_1} + \sqrt{\left(\frac{A_{ub}}{m_1}\right)^2+\frac{2}{m_1}\left(V_{\text{th}}\tau + B \right)},
\end{align}
where
\begin{align}
  \tau = \frac{C_{\text{eff}}}{G}, \quad B = \int_{-\infty}^{+\infty} \Bigg| \frac{W(\omega)}{\omega}  \Bigg| \,d\omega \nonumber.
\end{align}
\end{theorem}
The maximum comparator overdrive delay $t_{od}^{\text{max}}$  depends on the interference, hardware parameters, and design variable $\tau$.
The proof can be found in \cite{Avestruz2022a}.

The large\nobreakdash-signal transient performance of the power converter is determined by the inductor current slew rate, which in turn is determined by the minimum on time $T_{\text{on}}^{\text{min}}$. 
$T_{\text{on}}^{\text{min}}$ is a \emph{design objective} for comparator\nobreakdash-overdrive\nobreakdash-delay conditioning.
We design the comparator so that $T_{\text{on}}^{\text{min}}$ is as small as possible while maintaining stability.
Therefore, a judicious choice is to design the longest overdrive delay to equal the minimum on~time
% As long as the longest overdrive propagation delay does not exceed $T_{\text{on}}^{\text{min}}$, the current command can be adjusted so that the current control loop can stably operate at $T_{\text{on}}^{\text{min}}$. 
%The left hand side of (\ref{eqn:todmax_constraint}) is proven to be an upper bound of the overdrive delay. The detailed proof can be found in Appendix\,\ref{proof:gmax4stability}.
\begin{align}\label{eqn:todmax_tonmin}
    T_{\text{on}}^{\text{min}} = t_{od}^{\text{max}}. 
\end{align}

We define the \emph{normalized interference frequency} as
\begin{align} \label{eqn:def_omega_star}
    \hat{\omega} \triangleq \frac{\omega_{lb}}{\omega_b}, \quad \omega_b \triangleq \frac{2\pi}{T_{\text{on}}},
\end{align}
the \emph{normalized interference amplitude} as
\begin{align} \label{eqn:def_a_star}
    \hat{A} \triangleq \frac{A_{ub}}{A_b}, \quad  A_b \triangleq \frac{A_{ub}}{m_1T_{\text{on}}},
\end{align}
the \emph{normalized comparator time constant} as
\begin{align} \label{eqn:def_g_star}
    \hat{\tau} \triangleq \frac{\tau}{\tau_b}, \quad \tau_b \triangleq\frac{m_1T^2_{\text{on}}}{2V_{\text{trig}}},
\end{align}
and the \emph{normalized minimum on time} as
\begin{align} \label{eqn:def_on_time_star}
    \hat{T}_{\text{\text{on}}}^{\text{min}} \triangleq \frac{T_{\text{\text{on}}}^{\text{min}}}{T_{\text{\text{on}}}}.
\end{align}
The base value for the normalization were chosen to be the operating points in the steady state. For example, $T_{\text{on}}$ represents the on time of the current control loop in the steady state.

%As shown in Figs.~\ref{fig:copd_nomogram_p1},~\ref{fig:copd_nomogram_p2} and \ref{fig:copd_nomogram_p3}, the $x$\,-\,axis is the normalized interference amplitude and the $y$\,-\,axis is the normalized interference frequency.
%The contour line indicates different lower bound of the time constant of comparator $\hat{\tau}$.

% namely maximum gain $G^*$. We use the reciprocal gain because it is more linear with $A$ and $\omega$ than the gain, hence it makes the interpolation in the nomogram more accurate.

\begin{figure}
    \centering
    \includegraphics[width = 8 cm]{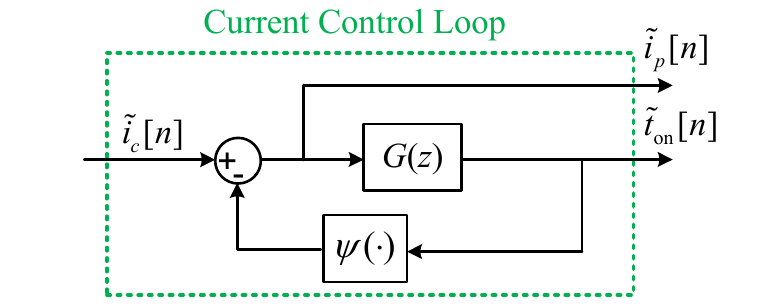} 
    \caption{The Lure system representation of current control loop with comparator overdrive for large\nobreakdash-signal analysis. The interference is embedded in $\psi(x)$.}
    \label{fig:luresystem_part2}
\end{figure}

We can acquire more intuition on the stability of the current control loop with comparator\nobreakdash-overdrive\nobreakdash-delay conditioning through linearization. 
The linearized model can be expressed as a block diagram in Fig.\;\ref{fig:luresystem_part2} with the linear feedback gain 
\begin{align} \label{eqn:S2}
\psi = \frac{f(T_{\text{on}}) - f(T_{fi})}{T_{\text{on}} - T_{fi}},
\end{align}
where $T_{fi}$ is the steady\nobreakdash-state value of (\ref{eqn:final_i_eqn_num_trick}) and $T_{\text{on}}$ is the on time in the steady state.

Interference manifests in the feedback gain $\psi$.
%as described in Section\,II-F of the Part I of the paper \cite{cmpartone2022}.
Negative $\psi$ results in positive feedback and may destabilize the current control loop.
Larger comparator time constant $\tau$ results in longer delay from $T_{fi}$ to $T_{\text{on}}$.
From (\ref{eqn:S2}), larger \mbox{$T_{\text{on}} - T_{fi}$} can decrease the amplitude of $\psi$ and stabilize the current control loop.  However, either positive or negative feedback can result in stable control loop. 

\begin{figure}
    \centering
    \includegraphics[width=8 cm]{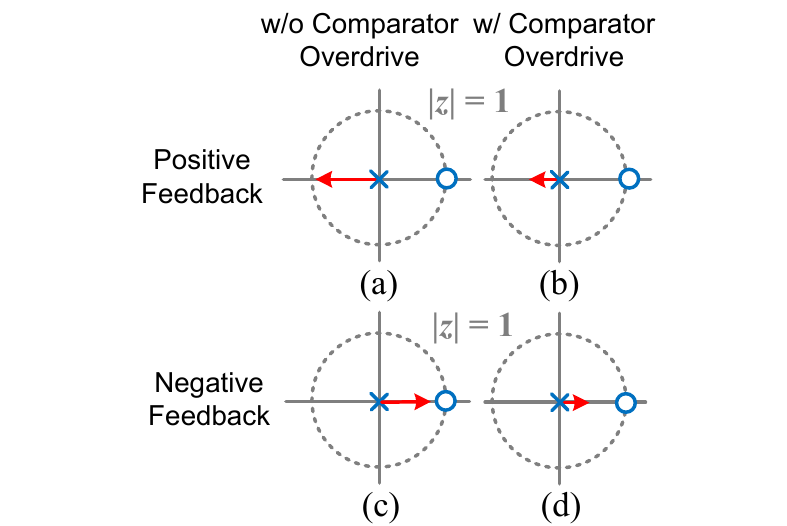}
    \caption{Small-signal root locus of the constant off(on)-time current control loop with the comparator overdrive delay. The sign of the feedback is determined by the interference. In positive feedback, comparator overdrive delay decreases the loop gain to increase the stability margin. In negative feedback, the closed\nobreakdash-loop pole is moved further to the left by filter.}
    \label{fig:lc_copd_cotcm}
\end{figure}

We can visualize the stabilizing effect of comparator\nobreakdash-overdrive\nobreakdash-delay conditioning through the root locus, which is shown in Fig.\,\ref{fig:lc_copd_cotcm}. We observe that comparator\nobreakdash-overdrive\nobreakdash-delay for positive feedback in Figs.\,\ref{fig:lc_copd_cotcm} (a) and (b) move the location of the worst\nobreakdash-case closed\nobreakdash-loop pole further to the right, hence improving stability margin.  Likewise, for negative feedback in Figs.\,\ref{fig:lc_copd_cotcm} (c) and (d), the closed\nobreakdash-loop pole moves further to the left.  In this way, comparator\nobreakdash-overdrive\nobreakdash-delay conditioning can guarantee that the closed\nobreakdash-loop poles stay inside the unit disk, hence guaranteeing stability.

We next show the quantitative relationship between interference and the location of closed\nobreakdash-loop pole. From (\ref{eqn:filter_lin_sys2}), the closed\nobreakdash-loop transfer function is 
\begin{align} % \label{eqn:iltf1} 
     C_2(z) = \frac{\beta }{1-az^{-1}},
\end{align}
where
\begin{align} %\label{eqn:filter_ztransform}
       \beta = \frac{m_1}{m_1 + \psi}, \quad a = \frac{\psi}{m_1 + \psi};
\end{align}
$\psi$ is defined in (\ref{eqn:S2}). Given an interference with amplitude upper bound $A_{ub}$ and frequency lower bound $\omega_l$, $\psi$ is bounded within the range $[\psi_{\text{min}}, \psi_{\text{max}}]$ where
\begin{align}
   \psi_{\text{min}} = \frac{-2m_1}{1 + \sqrt{1 + \frac{1}{\hat{A}^2}(\hat{\tau} - \frac{\hat{A}}{\hat{\omega}})}}, \nonumber \\
    \psi_{\text{max}} = \frac{2m_1}{-1 + \sqrt{1 + \frac{1}{\hat{A}^2}(\hat{\tau} - \frac{\hat{A}}{\hat{\omega}})}}.
\end{align}
The detailed derivations can be found in \cite{Avestruz2022a}.
For different operating points, the pole $a$ is always real and within the range \mbox{$[a_{\text{min}},a_{\text{max}}]$} where
\begin{align} \label{eqn:range_pole_sc}
a_{\text{min}} = \frac{\psi_{\text{min}}}{(m_1 + \psi_{\text{min}})}, \quad
a_{\text{max}} = \frac{\psi_{\text{max}}}{(m_1 + \psi_{\text{max}})}.
\end{align}
The condition for small\nobreakdash-signal stability is
\begin{align} \label{eqn:stab_pole_sc}
\abs{a_{\text{min}}}<1 , \quad \abs{a_{\text{max}}}<1.
\end{align}
The worst\nobreakdash-case settling and overshoot can be obtained from (\ref{eqn:settlecycle1_part2}) and (\ref{eqn:overshoot1_part2}), respectively. 
\begin{figure*}[ht]
\centering
\subfigure[The relationship between the worst\nobreakdash-case settling $N_w$ and the time constant $\hat{\tau}$ of the comparator. The frequency of the interference is $\hat{\omega} = 2$.]{
\includegraphics[width=0.3\textwidth]{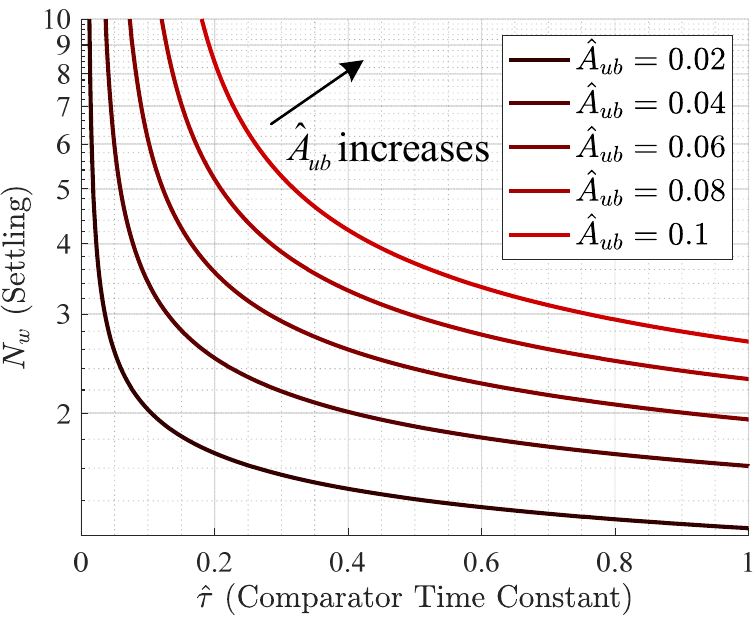} \label{fig:opdr_settlingtime}}
\subfigure[The relationship between the worst\nobreakdash-case overshoot $O_w$ and the time constant $\hat{\tau}$ of the comparator. The frequency of the interference is $\hat{\omega} \mbox{= 2}$.]{
    \includegraphics[width=0.3\textwidth]{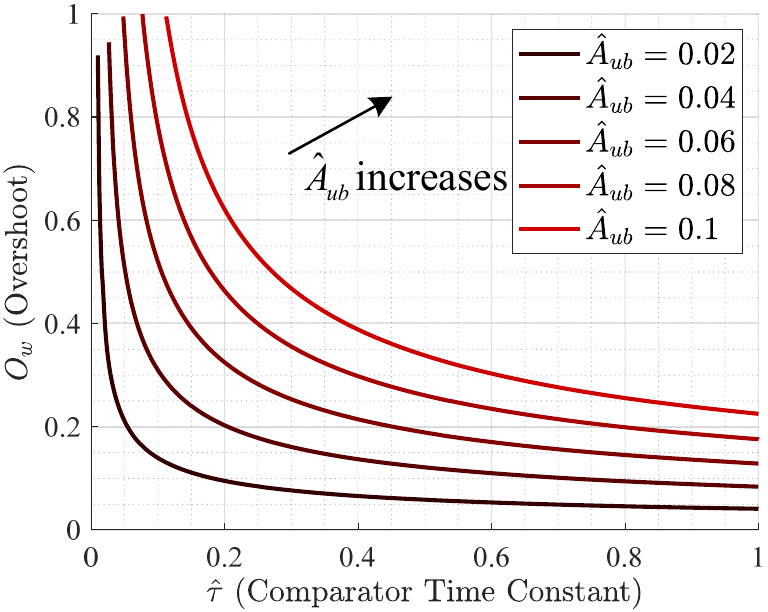}\label{fig:opdr_overshoot}}
\subfigure[The relationship between the maximum comparator overdrive delay $t_{od}^{\text{max}}$ and the time constant $\hat{\tau}$ of the comparator. The frequency of the interference is $\hat{\omega} = 2$.]{
    \includegraphics[width=0.3\textwidth]{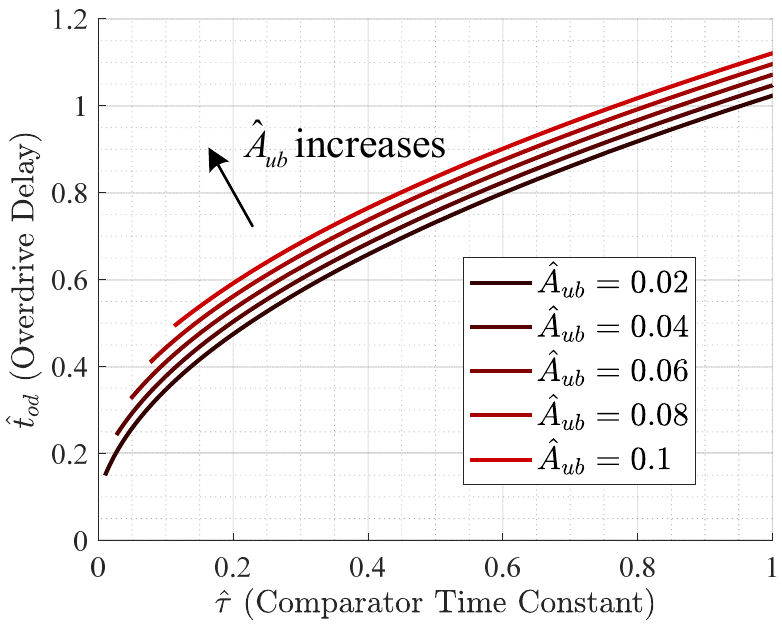} 
    \label{fig:opdr_overdrive_delay}
}
\caption{The design diagram for comparator overdrive given different interference levels.}
\label{fig:opdr_settling_overshoot}
\end{figure*}
Figures~\ref{fig:opdr_settlingtime}, \ref{fig:opdr_overshoot}, and \ref{fig:opdr_overdrive_delay} show how the settling, overshoot and overdrive delay change with the comparator time constant.
We observe that both the settling and overshoot decrease as comparator time constant $\hat{\tau}$ increases; also, overdrive delay increases as $\hat{\tau}$ increases. These observations match our intuition because the increase in the comparator time constant results in a corresponding increase in the ability of the comparator to attenuate interference; hence, overshoot and settling is smaller. However, with this increase in comparator time constant, overdrive delay is longer, in consequence sacrificing large\nobreakdash-signal speed.

The analysis in this section can be used for control conditioning fixed-frequency peak current-mode control by substituting $G(z)$ in (\ref{eqn:gzofcotcm_part2_filter}) by (\ref{eqn:fpccplanttf_part2}).
\begin{align} \label{eqn:fpccplanttf_part2}
    G(z) &= \frac{1-z^{-1}}{m_1 + m_2 z^{-1}},\\
    \psi(x) &= w(x + DT) - w(DT).
\end{align}
To extend the analysis to the constant on\nobreakdash-time control or the fixed\nobreakdash-frequency valley current\nobreakdash-mode control, we need to substitute $m_1$ by $m_2$ and $T_{\text{on}}$ by $T_{\text{off}}$ in  (\ref{eqn:def_omega_star}) and (\ref{eqn:def_a_star}).

In summary,  comparator\nobreakdash-overdrive\nobreakdash-delay conditioning can be a powerful control conditioning approach that can be easily implemented. The analog comparator, which is commonly used in peak current\nobreakdash-mode control, can directly have a dual\nobreakdash-use as an interference attenuator without extra complexity.
Within an integrated circuit, the level of attenuation can be easily chosen by adjusting the tail current.

\section{Hardware Results} \label{sec:hardwaredesignguide}

\begin{figure}
    \centering
    \includegraphics[width = 8cm]{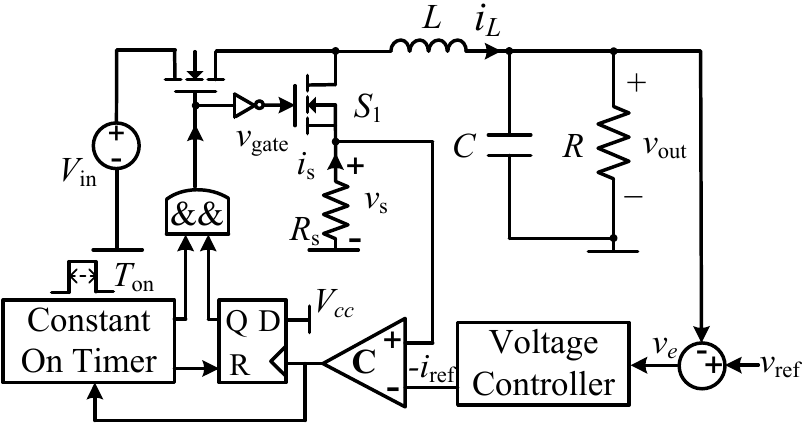}
    \caption{Schematic diagram of a digitally-controlled current-mode constant-on time buck converter.}
    \label{fig:schematicebuck}
\end{figure}

We use a multi\nobreakdash-megahertz buck converter prototype as a working hardware example to demonstrate the effectiveness of control conditioning for current control loops with interference dysfunction.  For this prototype, we selected constant on\nobreakdash-time current\nobreakdash-mode control, which is frequently used in high\nobreakdash-speed converters for microprocessors.
The schematic is shown in Fig.\,\ref{fig:schematicebuck}.

\subsection{DC-DC Converter Hardware Prototype Platform}
The conditioning methods in this paper apply for alleviating the effect of interference that occurs in all types of current sensors. We employ the ground\nobreakdash-referenced shunt\nobreakdash-resistor current sensor as a widely-used exemplar because of its simplicity and wide measurement bandwidth.
A 5\,MHz buck converter demonstrating cycle-by-cycle control is specified in Table\;\ref{table:cotcm_buck_param}.  This constant on-time current mode converter delivers 2\,V at 30\,Watts from a 12\,V input. 

\begin{figure}[htp]
        \centering
        \includegraphics[width = 7 cm]{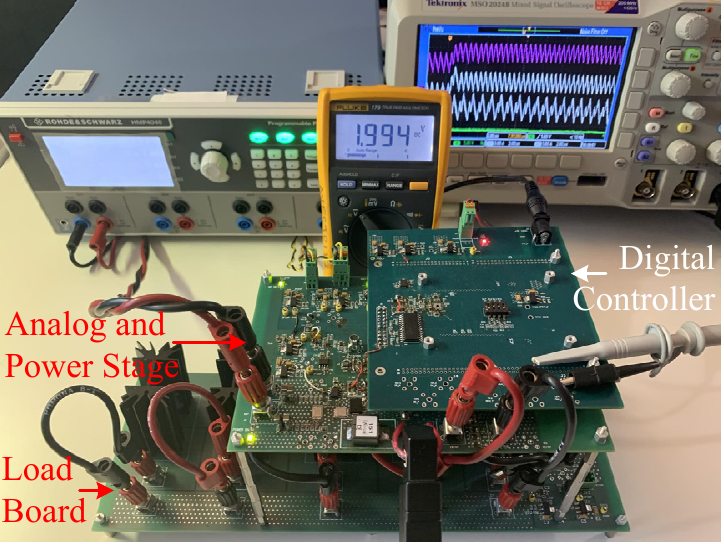}
        \caption{
       A 30\,W, 5\,MHz current\nobreakdash-mode buck converter using constant on\nobreakdash-time cycle\nobreakdash-by\nobreakdash-cycle digital control \cite{Cui2018a}. The digital controller is implemented in Xilinx Artix\nobreakdash-7 FPGA.
        }
        \label{fig:exp_ebuck_2}
\end{figure}

The control algorithms were implemented in a low cost Xilinx Artix-7 FPGA together with a
15\,MHz high-speed ADC and a 40\,MHz DAC are used in the digital system.
The FPGA communicates with the ADC through a 100\,MHz LVDS high-speed interface.
We implemented a hybrid digital and analog controller for current-mode control with commercial off-the-shelf comparators. The hardware is shown in Fig.\,\ref{fig:exp_ebuck_2}.
\begin{table}[tb]
    \caption{Design Parameters of the Constant On\nobreakdash-Time Current\nobreakdash-Mode Buck Converter}
    \label{table:cotcm_buck_param}
    \centering
    \begin{tabular}{cccccccccccccc}
        \toprule
\textbf{Parameters}&\textbf{Values}&\textbf{Parameters}&\textbf{Values}\\
\midrule
$\boldsymbol{V_{\textbf{in}}}$
& 12\,V & $\boldsymbol{L}$ & 240\,nH\\
\midrule
$\boldsymbol{V_{\textbf{out}}}$
& 2\,V & $\boldsymbol{R_{\textbf{load}}}$ & 0.2\,$\Omega$ \\
\midrule
$\boldsymbol{I_{\textbf{out}}}$ 
& 8\,A & $\boldsymbol{R_s}$ & 10\,m$\Omega$ \\
\midrule
$\boldsymbol{T_{\textbf{on}}}$ 
& 100\,ns & $\boldsymbol{f_{lb}}$
& 5\,MHz\\
\midrule
$\boldsymbol{C}$ & 100 $\mu$F & $\boldsymbol{v_{I}}$
& 4\,mV \\
%$\mathbf{T^{\textbf{min}}_{\textbf{off}}}$  & 175\,ns \\
        \bottomrule
    \end{tabular}
\end{table}

\subsection{Comparing the Three Control Conditioning Methods}
\label{sec:performancecomp}
% \begin{figure*}
%     \centering
%     \includegraphics[width=16cm]{designflow_cmc.pdf}
%     \caption{Design procedure for current mode control include two steps as shown in the top blue layer. current control loop design include three steps as shown in the middle red layer. Current conditioning design include three steps as shown in the bottom green layer.\red{change of to for and no capitalize/sovle the dynmaic mapping}}
%     \label{fig:designflow_cmc}
% \end{figure*}
\begin{figure}[htp]
\centering
\subfigure[Normalized amplitude $\hat{A}_{ub} = 0.01$, Normalized frequency $\hat{\omega}_{I} = 3$.]{
\includegraphics[width=7 cm]{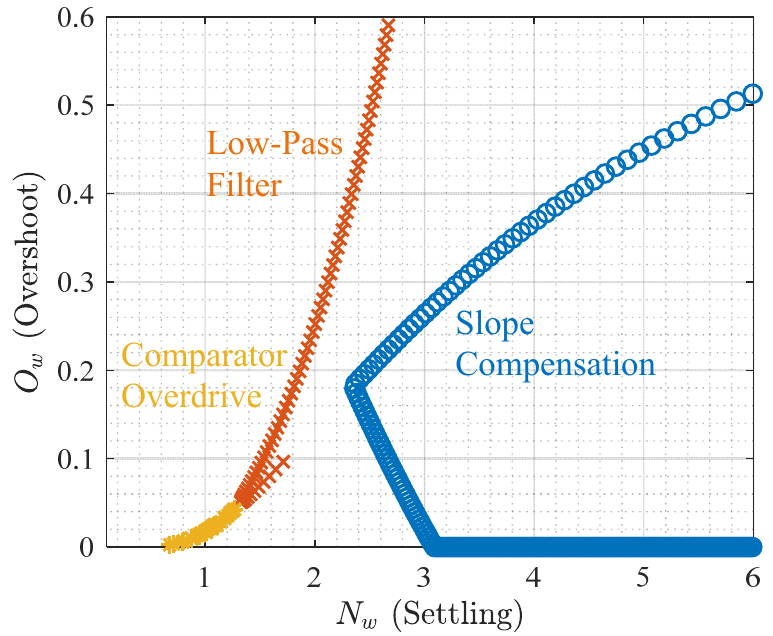}\label{fig:comparision_a_001_w3}
}
\newline
\subfigure[Normalized amplitude $\hat{A}_{ub} = 0.12$, Normalized frequency $\hat{\omega}_{I} = 1$.]{
    \includegraphics[width=7 cm]{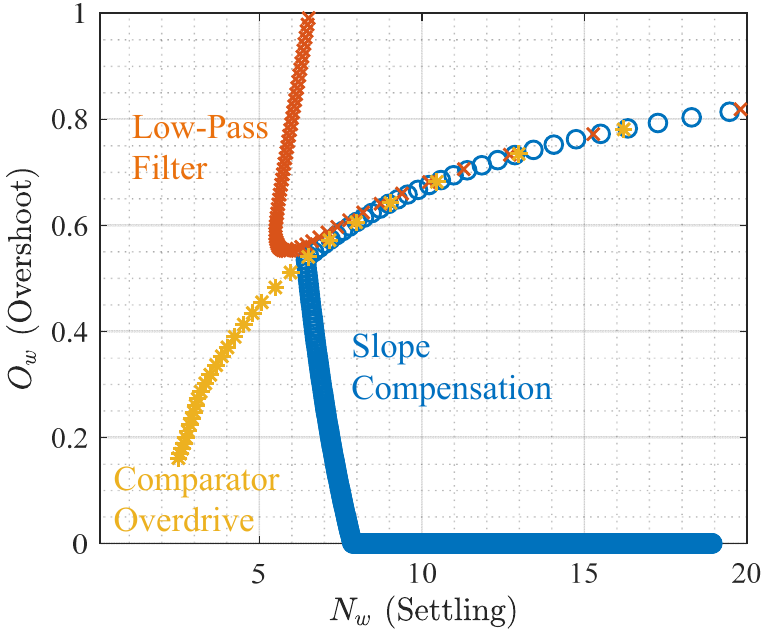} 
    \label{fig:comparision_a_012_w1}
}
\caption{Overshoot\nobreakdash-settling cycle performance tradeoff. Three interference conditioning methods are compared on a current\nobreakdash-mode converter using constant off\nobreakdash-time control. The $\hat{A}_{ub}$ and $\hat{\omega}_{I}$, which will be defined in Section \ref{subsection:opddesignguide}, are the normalized amplitude and frequency of interference.
% The $\tau^*$, which will be defined in Section \ref{subsection:lpf}, is the normalized time constant of filter.
}
%\label{fig:subfigureExample}
\end{figure}
Slope compensation, low-pass filtering, and comparator\nobreakdash-overdrive\nobreakdash-delay conditioning repair corrupted static and dynamical mappings. 
% We would like to compare the pros and Cons on different methods.
From a unified framework, we compare these three control conditioning methods.  All three methods decrease the degree of nonlinearity in the static mapping, but introduce other non\nobreakdash-idealities that must be addressed.  Low\nobreakdash-pass filtering repairs the nonlinearity caused by the interference, but introduces its own nonlinearity in the current sensor output; these two nonlinearities must be balanced to minimize the overall nonlinearity.
Slope compensation adds a gain error while 
comparator\nobreakdash-overdrive\nobreakdash-delay conditioning causes an offset error in the static mapping.  All three methods ensure stability and improve transient response of the current control loops.
To compare control\nobreakdash-conditioning methods, we use overshoot and settling as performance metrics.

It is inadequate to compare either the settling or overshoot in isolation because each method allows designers to trade\nobreakdash-off between these two performance metrics through \emph{design parameters}.  Instead, we compare among the tradeoff spaces and construct two\nobreakdash-dimensional performance spaces as shown in Figs.\,\ref{fig:comparision_a_001_w3} and \ref{fig:comparision_a_012_w1}. By varying the key design parameter for each method, e.g. compensation slope, filter time constant, or comparator time constant, we plot the point set for each of the three methods in this space. One observes that the points closer to the origin have a better performance tradeoff in overshoot and settling. These curves depend on the interference parameters $(A_{ub}, \omega_I)$, with all methods performing worse with higher interference amplitude. We define $\omega_I$ as the {\em{interference frequency band}}, which means the set of frequencies within which the interference frequency is contained. In this paper, the {\em{interference frequency}} refers to the interference frequency band.

Fig. \ref{fig:comparision_a_001_w3} illustrates that if switching frequency is much lower than the interference frequency (e.g., \mbox{$\omega_{sw} = \omega_{I}/3$)}, both the comparator\nobreakdash-overdrive\nobreakdash-delay conditioning and low\nobreakdash-pass filter achieve a smaller overshoot and shorter settling than slope compensation.
Fig. \ref{fig:comparision_a_012_w1} illustrates that if the switching frequency is comparable to the interference frequency (e.g., \mbox{$ \omega_{sw} \approx \omega_I$}), comparator overdrive\nobreakdash-delay\nobreakdash-conditioning and slope compensation result in a smaller overshoot and shorter settling than the low\nobreakdash-pass filter.
For comparator\nobreakdash-overdrive\nobreakdash-delay conditioning, both the overshoot and settling monotonically increase together.  However, because the reduction of overshoot and settling is effected by increasing the comparator overdrive delay, the minimum on time is also increased. It is worth noting that this is so because the minimum on time is equal to comparator overdrive delay, i.e. the decision to turn the switch off cannot be made before the comparator can output its comparison. This minimum on time saturation has consequences in the large signal design of the overall power converter control loop. A qualitative comparison of the three conditioning methods is summarized in Table\,\ref{table:cond_compare}.

\begin{comment}
% \begin{table*}[tb]
%     \caption{Comparison of Three Conditioning Methods}
%     \label{table:cond_compare}
%     \centering
%     \begin{tabular}{cccccccccccccc}
%   \toprule
% \textbf{Method}
% &\textbf{Less} &\textbf{Large\nobreakdash-Signal}
% &\textbf{Transient Performance}
% &\textbf{Transient Performance}
% \\
% \textbf{}
% &\textbf{Complexity} &\textbf{Design Performance}
% &\textbf{@ $\omega \gg \omega_{\text{int}}$}
% &\textbf{@ $\omega \approx \omega_{\text{int}}$}
% \\
% \midrule
% \textbf{Filter}
% & +++ & + & ++ &  + \\
% \midrule
% \textbf{Slope Compensation} 
% & ++ & +++ & + & ++ & \\
% \midrule
% \textbf{Comparator Overdrive}  
% & + & ++
% & +++ & +++ \\
% %$\mathbf{T^{\textbf{min}}_{\textbf{off}}}$  & 175\,ns \\
%         \bottomrule
%     \end{tabular}
% \end{table*}
\end{comment}
\begin{table*}[tb]
    \caption{Comparison of Three Conditioning Methods}
    \label{table:cond_compare}
    \centering
    \begin{tabular}{cccccccccccccc}
  \toprule
\textbf{Method}
&\textbf{Less}
&\textbf{Transient Performance}
&\textbf{Transient Performance}
\\
\textbf{}
&\textbf{Complexity}
&\textbf{@ $\omega_s \ll \omega_I$}
&\textbf{@ $\omega_s \approx \omega_I$}
\\
\midrule
\textbf{Filter}
& +++ & ++ &  + \\
\midrule
\textbf{Slope Compensation} 
& ++  & + & ++ & \\
\midrule
\textbf{Comparator Overdrive}  
& +
& +++ & +++ \\
%$\mathbf{T^{\textbf{min}}_{\textbf{off}}}$  & 175\,ns \\
        \bottomrule
    \end{tabular}
\end{table*}

The transient performance of the filter is the best when the switching frequency is well below the interference frequency.  When the switching frequency is within the range of the interference, both slope compensation and comparator-overdrive-delay conditioning perform well. 
However, comparator-overdrive-delay conditioning incurs a higher complexity when implemented discretely, but is much more straightforward within an integrated circuit; additionally, as mentioned above, the dependence on minimum on time adds an additional constraint.  Adaptive tuning is straightforward for both slope compensation and comparator-overdrive-delay compensation.

\subsection{Control Conditioning Using Slope Compensation}
\begin{figure}[htp]
    \centering
    \includegraphics[width = 8cm]{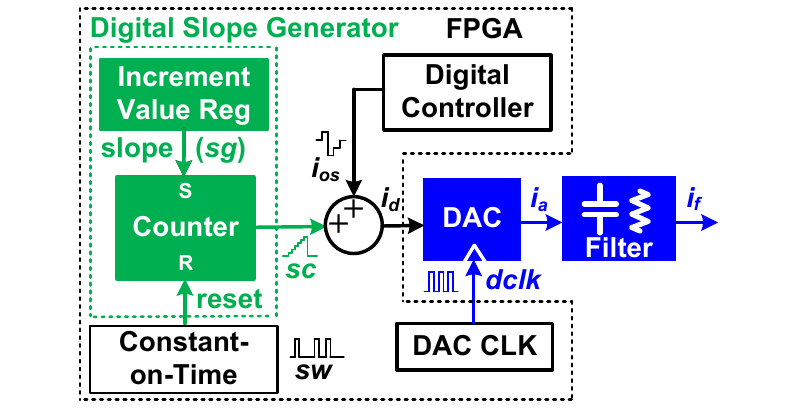}
    \caption{Schematic of the programmable digital slope generator, which allows for re-programmability and adaptability.}
    \label{fig:impl_sch_sc}
\end{figure}

\begin{figure}
    \centering
    \includegraphics[width = 8cm]{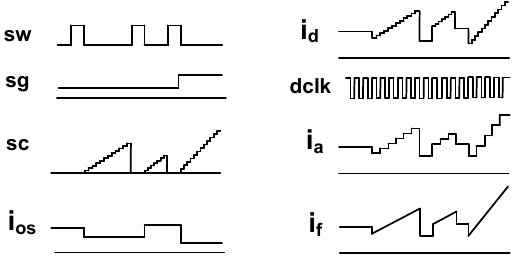}
    \caption{Waveform of the programmable digital slope generator shows straightforward implementation of digital functions.}
    \label{fig:impl_sch_waveform}
\end{figure}

We exhibit a digital slope generator with programmable slope in Fig.\,\ref{fig:impl_sch_sc}. The slope is programmed by the increment value register $\bm{sg}$. The output of the digital controller $\bm{i_{os}}$ determines the offset value of the valley current. The counter $\bm{sc}$ is triggered and reset every cycle. The constant-on-time modulator $\bm{sw}$ resets the counter to 0 at each rising edge and freezes the counter during the entire on time.
The counter starts at the falling edge of the constant-on-time modulator $\bm{sw}$ and counts upwards. $\bm{i_d}$ represents the summation of the valley current offset $\bm{i_{os}}$ and slope $\bm{sc}$. $\bm{i_d}$ is converted into an analog signal by the DAC. The output of the DAC $i_a$ is updated at the rising edge of the DAC clock. The filter smooths the stepped $i_a$ so that it is nearly an ideal ramp.
The digital slope generator and controller are implemented in the FPGA, and the DAC and filter are implemented in discrete hardware. This all-digital slope generator can be reconfigured in real-time to optimize for transient and stability at different operating points, in which gain scheduling is an example. All the critical waveforms can be found in Fig.\,\ref{fig:impl_sch_waveform}.

%The digital slope generator is verified in hardware.

\begin{figure}[htp]
\centering
    \subfigure[Steady state.]
    {
    \centering
    \includegraphics[width = 7cm]{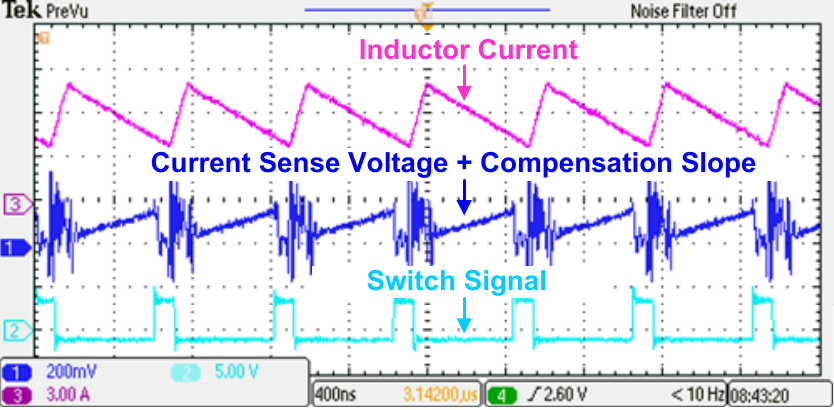}
    \label{fig:sc_steady_10}
    }
    \subfigure[Transient.]
    {
    \centering
    \includegraphics[width = 7cm]{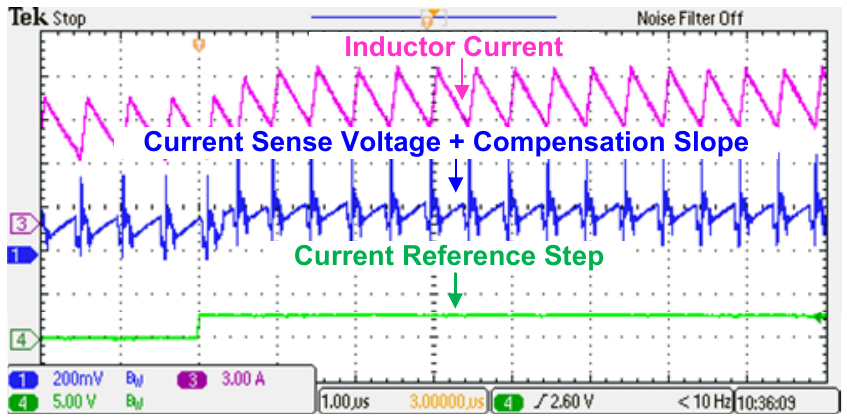}
    \label{fig:sc_trans_10}
    }
    \caption{Experimental waveforms of digital slope generator with slope \mbox{= 10\,A/$\mu$s} shows stable power converter operation despite a current controller corrupted by interference.}
\end{figure}

The output of the slope generator can function correctly both in steady state and in transition as shown in Figs.\,\ref{fig:sc_steady_10} and \ref{fig:sc_trans_10}. The slope generator output is shown in Channel 1 (dark blue). The inductor current is shown in  channel 3 (pink)  and the switching signal is shown in channel 2 (light blue). There is a propagation delay from the switching signal to the inductor current. We observe that the slope generator output is highly contaminated by the switch-on transient, but the slope compensation still functions correctly because we only need to compensate the current sensor slope during the off time, which means the switch-off transition does not induce heavy interference on the slope generator output.

\begin{figure}[htp]
\centering
    \subfigure[Slope compensation = 0\,A/$\mu$s.]{
    \centering
    \includegraphics[width = 7cm]{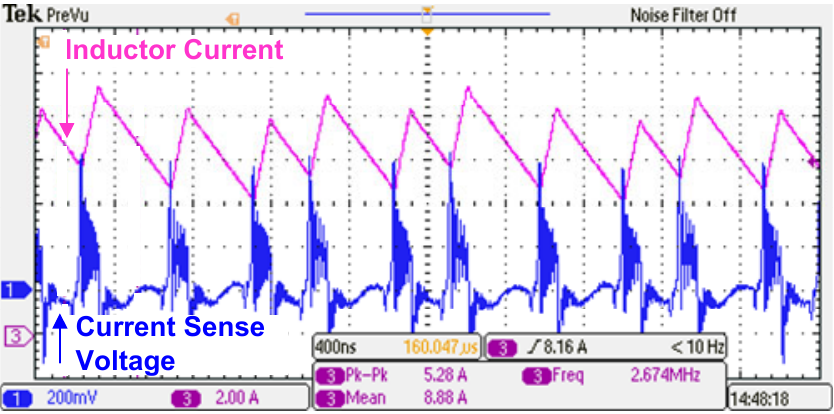}
    \label{fig:nonsc_zoomin}
    }
    \subfigure[Slope compensation = 10\,A/$\mu$s.]{\centering
    \includegraphics[width = 7cm]{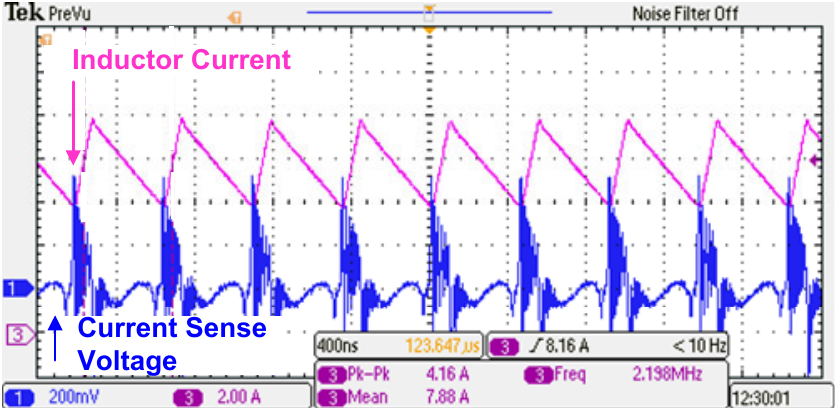}
    \label{fig:sc_zoomin}
    }
    \caption{Inductor current and current-sense voltage of a current control loop corrupted by interference shows instability without slope compensation and stability.}
\end{figure}

\begin{figure}[htp]
\centering
    \subfigure[Slope compensation = 0\,A/$\mu$s.]
    {
    \centering
    \includegraphics[width=7 cm]{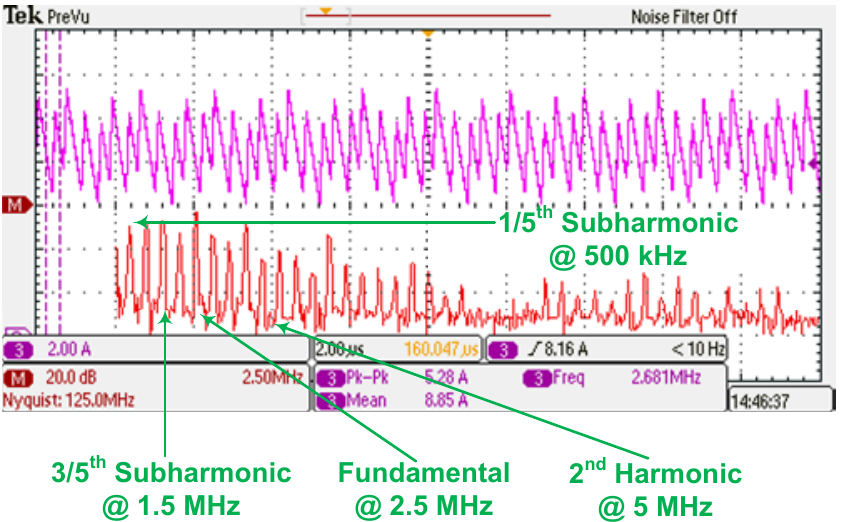}
    \label{fig:nosc_fft}
    }
    \subfigure[Slope compensation = 10 A/$\mu$s.]
    {
    \centering
    \includegraphics[width=7cm]{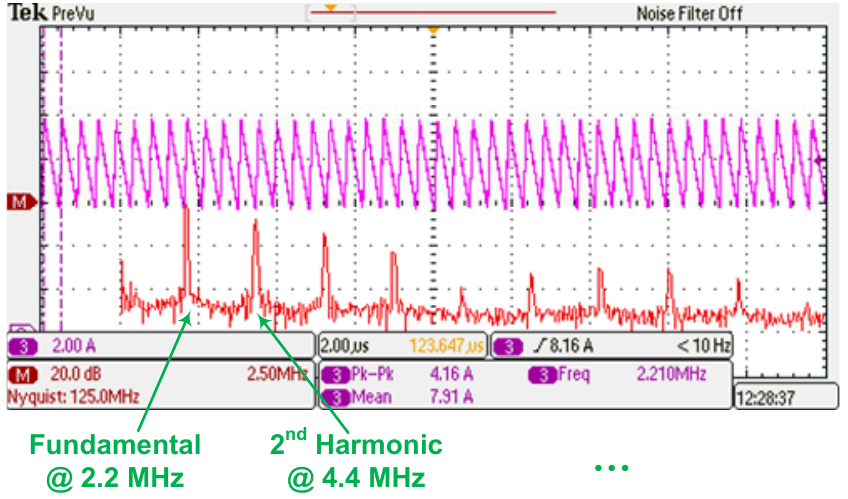}
    \label{fig:sc_fft}
    }
    \caption{Fourier transform of the inductor current of the current controllers that corrupted by interference show subharmonics without slope compensation, but stable harmonics with appropriate compensation slope.}
\end{figure}

Without slope compensation, the inductor current is not stable because of the contaminated current sense voltage as illustrated in Fig.\,\ref{fig:nonsc_zoomin}.
The subharmonic frequencies of one\nobreakdash-fifth to fourth-fifths of the fundamental frequency appear in Fig.\,\ref{fig:nosc_fft}.
With slope compensation of 10\,A$/\mu$s, the inductor current is stabilized despite the severe contamination on the current sensor voltage as shown in Fig.\,\ref{fig:sc_zoomin}.
The only remaining harmonics are positive integer multiples of the fundamental frequency as validated by the Fourier spectrum in Fig.\,\ref{fig:sc_fft}.
%The steady-state inductor currents with and without the slope compensation are compared in Figs.\,\ref{fig:nonsc_zoomin} and \ref{fig:sc_zoomin}.
%The Fourier transform of the inductor current waveforms in Figs.\,\ref{fig:nosc_fft} and \ref{fig:sc_fft} are shown in Figs.\,\ref{fig:nonsc_zoomin} and \ref{fig:sc_zoomin}, respectively.
%The steady-state frequency in Figs.\,\ref{fig:nosc_fft} and \ref{fig:sc_fft} are slightly different because of the slight difference in output voltage. The hardware result verifies the theory that the inductor current can be stablized by a compensation slope.
% Note that Figure.\,\ref{fig:nonsc_zoomin} and Fig.\,\ref{fig:nosc_fft} are the same as Fig.\,\ref{fig:filterss_4_7p_zoomin} and Fig.\,\ref{fig:filterss_4_7p_zoomout} because they both are control group. 
%From the scope plot, it is not easy to capture the dynamic response of the current control loop. We export the CSV file and plot in using Matlab for better illustration. 
%The current control loop is unstable in hardware when the compensation slope is 0\,A/$\mu$s in Fig. xxx. When the slope is increased to 10\,A/$\mu$s, 
%which is closed to the design value in (\ref{eqn:act_opt_slope}). 
%the current controller settles quickly to a stable equilibrium within several cycles in Fig. xxx.
%The ringing disappears when the slope is further increased to \,A/$\mu$s. However, the current transient takes longer to settle. The transient response of the hardware agrees with theory in that there is an optimal compensation slope to optimize settling and overshoot.

\begin{figure}[htp]
\centering
    \subfigure[Experimental waveforms from oscilloscope.]{
    \centering
    \includegraphics[width = 8cm]{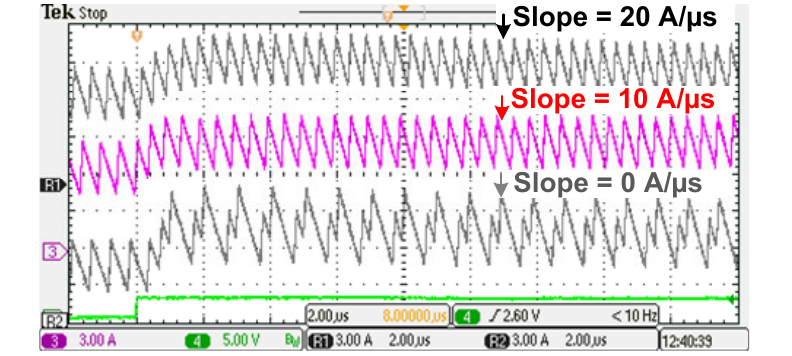}
    \label{fig:slopetransexp}
    }
    \subfigure[Experimental waveforms with the inductor current valleys marked as discrete time samples.]{\centering
    \includegraphics[width = 8cm]{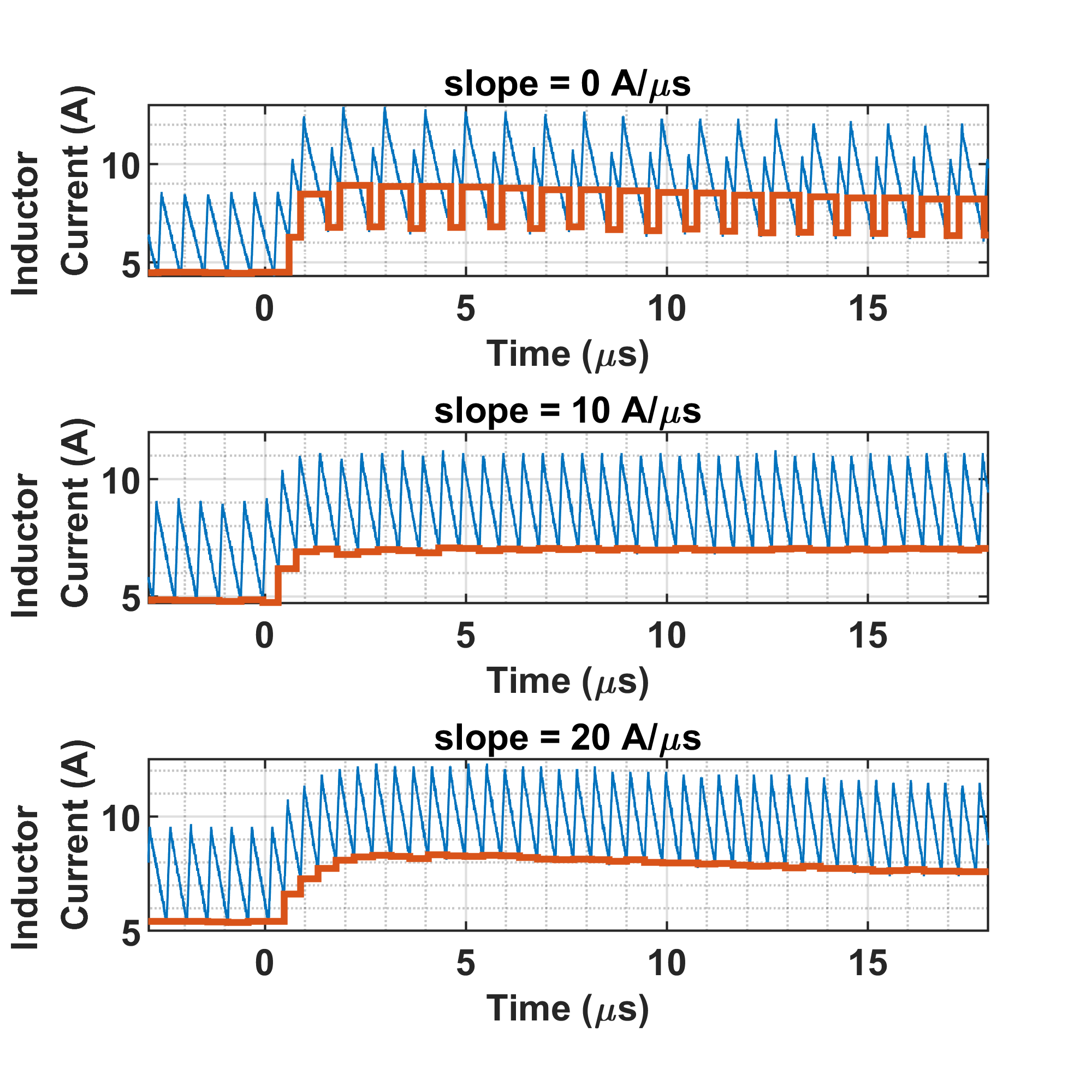}
    \label{fig:slopetranssimu}
    }
    \caption{Experimental step-up response of the current control loop with different slope compensations shows instability with no slope compensation, stability with appropriate compensation, and slow transient response with over-compensation. }
\end{figure}

The slope compensation results in a second-order response to the current control loop as illustrated in Fig.\,\ref{fig:slopetransexp}.
The greater the compensation slope, the slower the transient response as shown in Fig.\,\ref{fig:slopetranssimu}.
The current control loop is unstable in hardware when the compensation slope is 0\,A/$\mu$s. When the slope is increased to 10\,A/$\mu$s, 
%which is closed to the design value in (\ref{eqn:act_opt_slope}). 
the current controller settles quickly to a stable equilibrium within several cycles.
The ringing disappears when the slope is further increased to 20\,A/$\mu$s. However, the current transient takes longer to settle.

%The current controller goes unstable with zero slope compensation. The current controller shows slow transient response if it is over-compensated by a slope compensation of 20\,A/$\mu$s.

% inductor current waveform
% FFT of the inductor
% Transient waveform
% \begin{figure}[htp]
%     \centering
%     \includegraphics[width = 6cm]{Figure/slopetransexp.pdf}
%     \caption{Experimental step-up response of the current controller with different slope compensations shows instability with no slope compensation, stability with appropriate compensation, and slow transient response with over-compensation. }
%     \label{fig:slopetransexp}
% \end{figure}
% \begin{figure}[htp]
%     \centering
%     \includegraphics[width =6cm]{Figure/slopetranssimu.pdf}
%     \caption{Hardware step-up response of the current controller with different slope compensations. The current controller goes unstable with zero slope compensation. The current controller is stabilized by a compensation slope of 5\,A/$\mu$s. The current controller shows slow transient response if it is over-compensated by a slope compensation of 20\,A/$\mu$s.}
%     \label{fig:slopetranssimu}
% \end{figure}

\subsection{Control Conditioning Using Low-Pass Filter}
We use the first\nobreakdash-order low\nobreakdash-pass filter for illustration.
The design of the low\nobreakdash-pass filter includes the design of time constant $\tau$.
% We first find the range of $\tau$ to guarantee the global asymptotic stability by Theorem \ref{theorem:taomin4stability}.
% Within this range, we next design for transient performance using the transient performance design diagram in Fig.\,\ref{fig:no_tradeoff_lpf}.
% A practical design objective is to minimize settling time given an upper bound on the overshoot.
% Depending on the interference and overshoot constraint, there might be two types of design patterns.
% In pattern $A$, the overshoot constraint is loose, hence the minimal settling time can be achieved by varying the filter time constant.
% In pattern $B$, the overshoot constraint is tight, hence the settling time is sacrificed.
\begin{figure}[htp]
    \subfigure[$C$= 4.7\,pF,\,$R$ =  730\,$\Omega$]{
    \centering
    \includegraphics[width = 7 cm]{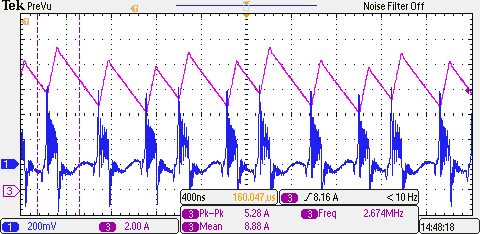}
    \label{fig:filterss_4_7p_zoomin}
    }
    \centering
    \subfigure[$C$= 22\,pF,\,$R$ =  730\,$\Omega$]{
    \centering
    \includegraphics[width = 7 cm]{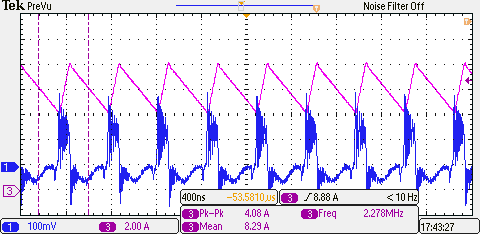}
    \label{fig:filterss_22p_zoomin}
    }
    \centering
    \subfigure[$C$= 68\,pF,\,$R$ =  730\,$\Omega$]{
    \centering
    \includegraphics[width = 7 cm]{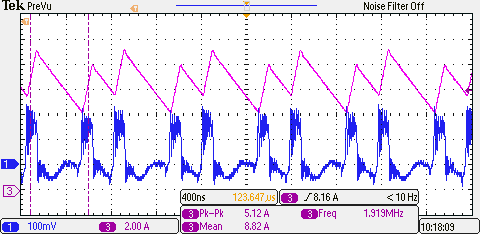}
    \label{fig:filterss_47p_zoomin}
    }
    \caption{Inductor current and current\nobreakdash-sense output of the current control loops with different filters.}
\end{figure}
\begin{figure}[htp]
    \subfigure[$C$= 4.7\,pF,\,$R$ =  730\,$\Omega$]{
    \centering
    \includegraphics[width = 7 cm]{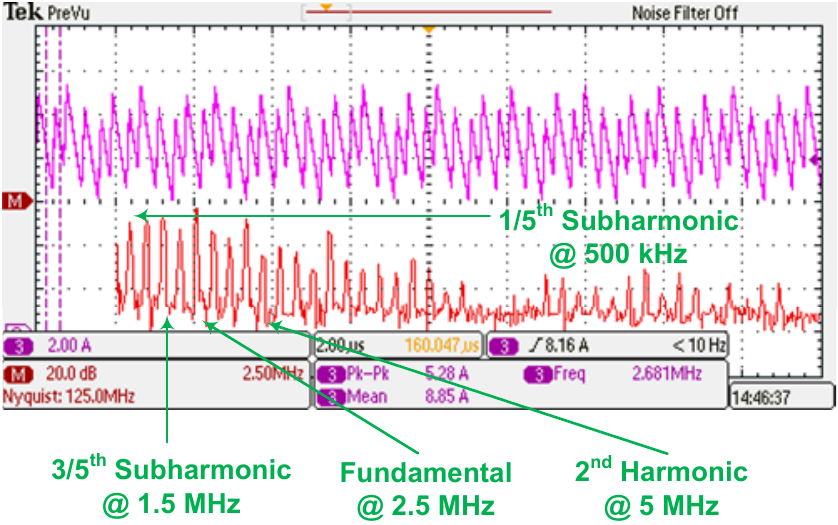}
    \label{fig:filterss_4_7p_zoomout}
    }
    \centering
    \subfigure[$C$= 22\,pF,\,$R$ =  730\,$\Omega$]{
    \centering
    \includegraphics[width = 7 cm]{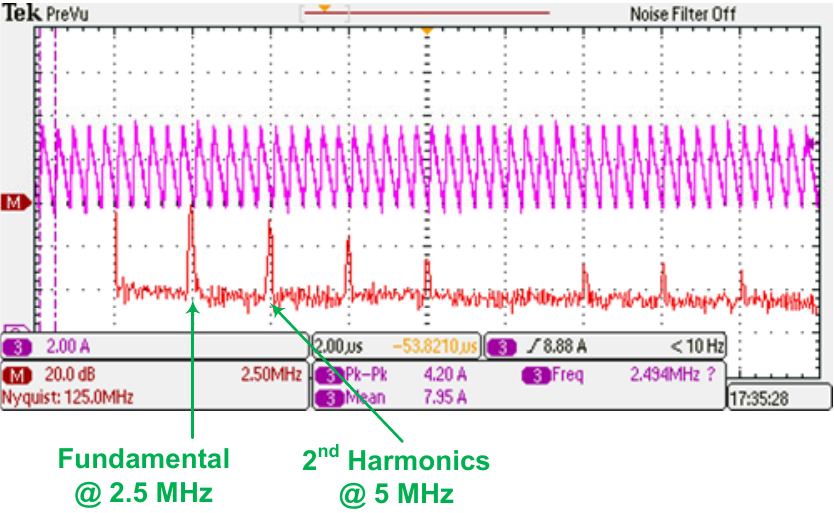}
    \label{fig:filterss_22p_zoomout}
    }
    \centering
    \subfigure[$C$= 68\,pF,\,$R$ =  730\,$\Omega$]{
    \centering
    \includegraphics[width = 7 cm]{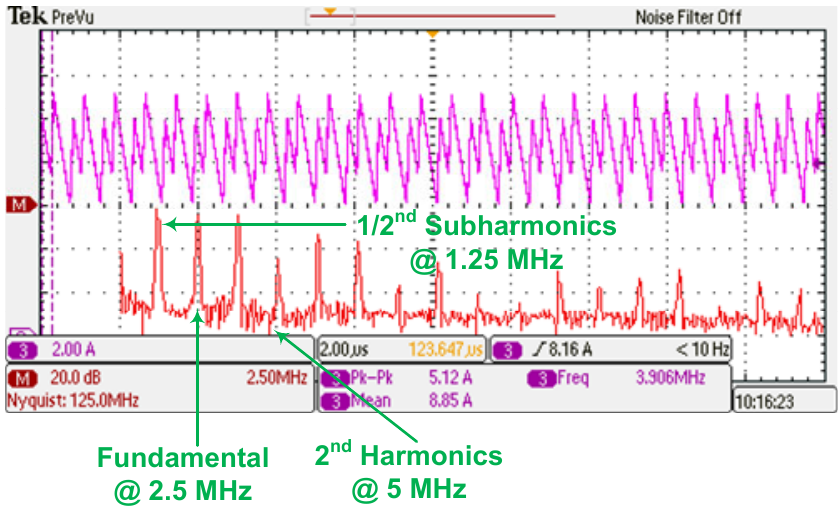}
    \label{fig:filterss_47p_zoomout}
    }
    \caption{Fourier transform of the inductor current of the current control loops with different filters.}
\end{figure}
\begin{figure}[htp]
    \centering
    \includegraphics[width = 7 cm]{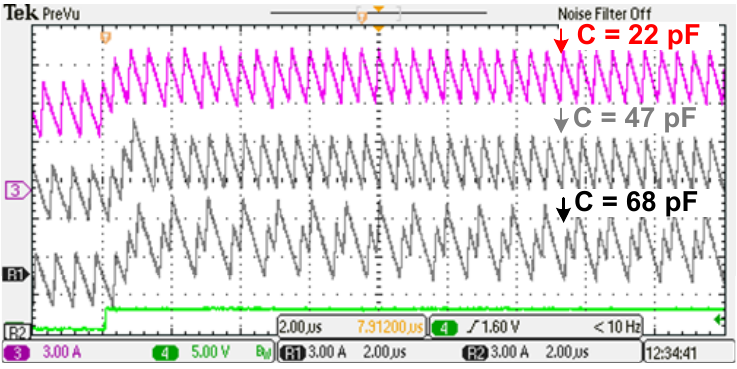}
    \caption{Experimental step\nobreakdash-up response of the current control loop with different filters.}
    \label{fig:filtertransexp}
\end{figure}
\begin{figure}[htp]
    \centering
    \includegraphics[width = 8 cm]{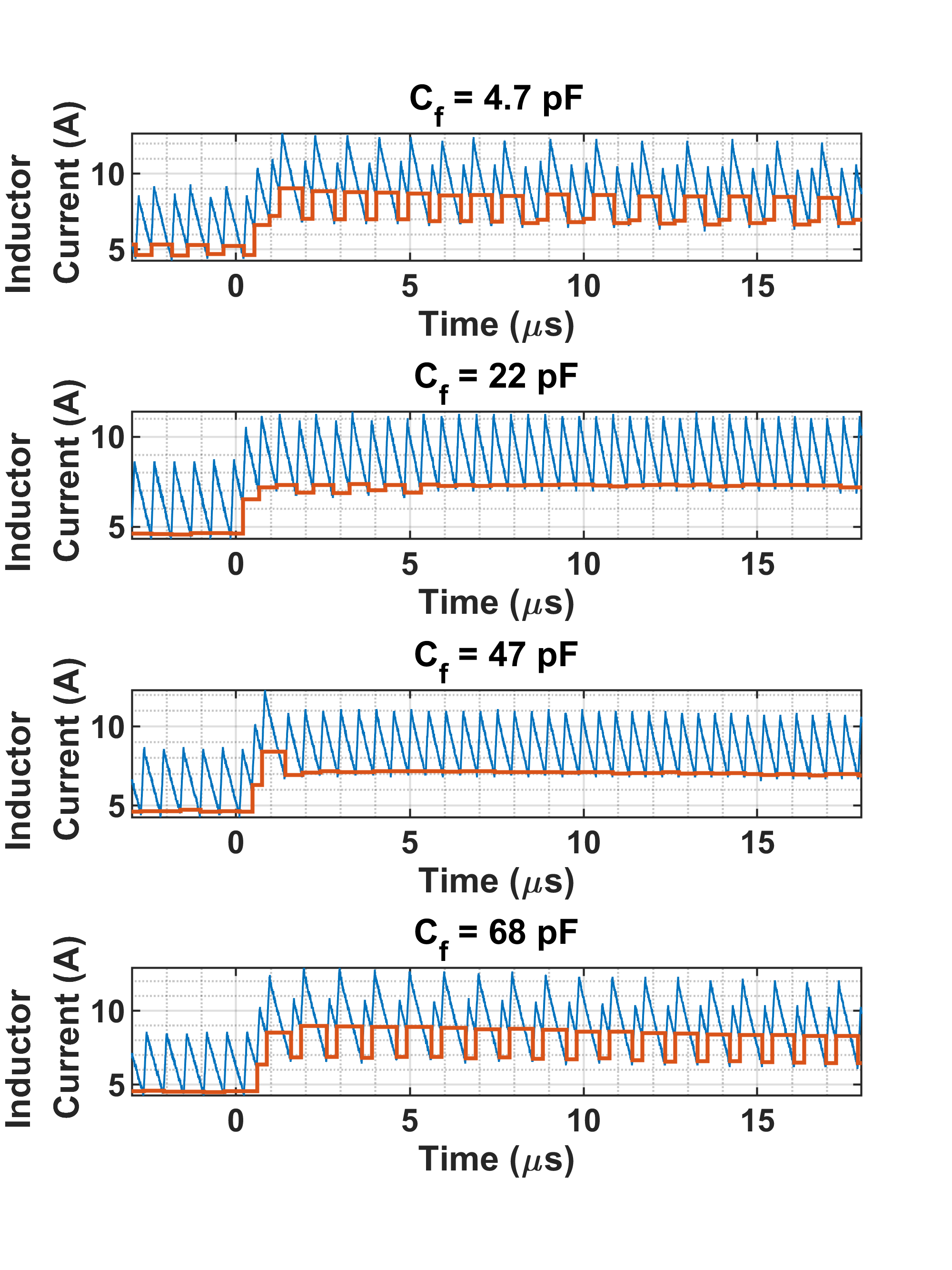}
    \caption{Step\nobreakdash-up response of the current control loop with different filters.}
    \label{fig:filtertranssimu}
\end{figure}
% Assume the interference is characterized by $(\hat{A}_{ub}, \hat{\omega}_{lb}) = (0.03, 2)$. If it is required that the upper bound of overshoot is $20\%$, as shown in the case $A$ in Fig.\,\ref{fig:no_tradeoff_lpf}, the minimal settling time occurs at the green square on the performance tradeoff curve. 
% This is also the global minimal settling time.
% The filter time constant $\hat{\tau}^*$ can be obtained at the minimum of the filter time constant curve --- settling time curve parameterized by $\hat{A}_{ub} = 0.03$, as shown in Fig.\,\ref{fig:filter_settlingtime}.
% If it is required that the constraint of overshoot is $10\,\%$ as shown in the case $B$ in Fig.\,\ref{fig:no_tradeoff_lpf}, the design constraints cannot be satisfied by only using low\nobreakdash-pass filter.
% The minimal settling time occurs at the green dot which is located at the crossing of the performance trade-off curve and overshoot constraint.
% Here we interpret (\ref{eqn:timescaleassumption2}) as
% \begin{align} \label{eqn:timescaleassumption3}
%     \omega^*_{\text{int}}\;c \ge 3.
% \end{align}
The constant on\nobreakdash-time buck prototype is used as a hardware design example.
% The range of $\hat{\tau}$ can be found through Theorem \ref{theorem:taomin4stability}.
% Given $(\hat{A}_{ub}, \hat{\omega}_{lb}) = (0.1, 2.5)$, to guarantee the global asymptotic stability, the normalized filter time constant $\hat{\tau}$ satisfies 
% \begin{align}
%     \hat{\tau} \ge 0.032.
% \end{align}
% The actual filter time constant $\hat{\tau}$ satisfies 
% \begin{align}
%     \tau = \hat{\tau} \times T_{\text{off}} \ge 16\,\text{ns}.
% \end{align}
% The design objective to minimize the settling with small enough overshoot. From Fig.\,\ref{fig:filter_settlingtime} and Fig.\,\ref{fig:filter_overshoot}, we choose $\hat{\tau} = 0.5$.
% We finally verify that the selected $\tau^*$ satisfies the assumption (\ref{eqn:uplimitonc}) and (\ref{eqn:timescaleassumption}).
% The assumption (\ref{eqn:timescaleassumption}) can be equivalently expressed as
% \begin{align} \label{eqn:timescaleassumption2}
%     \omega^*\;\tau^* \ge 1.
% \end{align}
% We next verify if the product of $\omega^*$ and $\tau^*$ satisfies (\ref{eqn:timescaleassumption2}). 
% In this example, given $\hat{\omega} = 2.5$, $\hat{\omega}\, \hat{\tau} = 1.25 \ge 1$ satisfies the condition.
% Therefore, the time constant of the low\nobreakdash-pass filter is designed as $\tau = 0.5\,T_{\text{on}}$.
Figures \ref{fig:filterss_4_7p_zoomin}, \ref{fig:filterss_22p_zoomin}, and \ref{fig:filterss_47p_zoomin} show the experimental waveforms of the current\nobreakdash-mode buck converter with different conditioning filters. We vary the capacitor value to change the time constant of the first\nobreakdash-order low\nobreakdash-pass filter. The inductor current is shown in Channel 3 (pink) and the current sensor output is shown in Channel 1 (dark blue), which is highly contaminated by interference. Note that the filter capacitor \mbox{$C=4.7$\,pF} is always placed at the comparator input, not only for low\nobreakdash-pass filter conditioning, but also for slope compensation and comparator-overdrive-delay conditioning because it can condition the voltage spike during switching transients, in other words, limit the bandwidth of the current\nobreakdash-sense output. This design not only protects the comparator from being overdriven, but also makes the assumption that the bandwidth limit holds in Definition 2 in the Part I of this paper \cite{cmpartone2022}.

If only the bandwidth limiting filter is used without proper design, the converter cannot function properly because inductor current exhibits a complicated subharmonic behavior as shown in Fig.\,\ref{fig:filterss_4_7p_zoomin}. From the Fourier transform of inductor current in Fig.\,\ref{fig:filterss_4_7p_zoomout}, we 
observe that in addition to the switching frequency component, the $1/5^{\text{th}}$, $2/5^{\text{th}}$, $3/5^{\text{th}}$, and $4/5^{\text{th}}$\nobreakdash-order subharmonics are also commingled because of the interference.
We increase the time constant of low\nobreakdash-pass filter by increasing the capacitance to 22\,pF. As shown in Fig.\,\ref{fig:filterss_22p_zoomin}, the inductor current goes back to stable periodic steady state; Fig.\,\ref{fig:filterss_22p_zoomout}, a longer\nobreakdash-horizon view, further confirms our observation. The Fourier transform of inductor current only contains switching\nobreakdash-frequency and higher\nobreakdash-order harmonics. A further step to increase the capacitance to 68\,pF destabilizes the inductor current as shown in Fig.\,\ref{fig:filterss_47p_zoomin}. Figure\,\ref{fig:filterss_47p_zoomout} shows that unlike Fig.\,\ref{fig:filterss_4_7p_zoomin}, the unstable inductor current in Fig.\,\ref{fig:filterss_47p_zoomin} only contains the $1/2^{\text{th}}$ order subharmonics. Experimentally, the steady state of the current control loop matches the theory. A fast filter cannot condition the interference and hence cannot decrease the nonlinearity of the static current mapping; a slow filter can distort the linear ramp and increase the nonlinearity.

The low\nobreakdash-pass filter affects the transient response of the current control loop. We test the step\nobreakdash-up response of the current control loop using four different low pass filters for control conditioning as shown in Fig.\,\ref{fig:filtertransexp}.  For better illustration, we export the oscilloscope data to Matlab and plot them in Fig.\,\ref{fig:filtertranssimu}. The valley currents are traced as a step plot (in red) to emphasize that we are directly controlling the valley current in this buck converter using constant\nobreakdash-on time control. Using \mbox{$C_f = 4.7$\,pF} results in an unstable steady state; however, the current control loop is stabilized when the filter capacitor is increased to \mbox{$C_f = 22$\,pF}. The tradeoff to stabilizing the loop is the incurrence of 10 cycles of settling. The settling transient is faster when the filter capacitor is increased to \mbox{$C_f = 47$\,pF}; however, there is a relatively large overshoot during transients, which can be explained by the zero in (\ref{eqn:iltf1}) introduced by the filter. When \mbox{$C_f = 68$\,pF}, the current control loop devolves into a 2\nobreakdash-cycle subharmonic instability.
%The experimental transition of current control loop matches the theory that there exist optimal filter time constant to optimize the settling and overshoot.

% step 3.3
\subsection{Control Conditioning Using Comparator\nobreakdash-Overdrive\nobreakdash-Delay} \label{subsection:opddesignguide}

\begin{comment}
\begin{figure*}[t]
\centering
\subfigure[Interference frequency \text{$\hat{\omega} \in [0.1,1]$}
]{\includegraphics[width=0.3\textwidth]{nomogram_p1_c.pdf} 
\label{fig:copd_nomogram_p1}
%\caption{Interference frequency }
}
\subfigure[Interference frequency \text{$\hat{\omega} \in [1,3]$}
\newline]{
    \includegraphics[width=0.3\textwidth]{    design_comp.pdf}
    %\rule{4cm}{3cm}
   \label{fig:copd_nomogram_p2}
   %\caption{Interference frequency $\omega_{\text{int}} \in [0.5,3]$}
}
\subfigure[Interference frequency \text{$\hat{\omega} \in [0.1,3]$}]{
    \includegraphics[width=0.3\textwidth]{nomogram_p3_c.pdf}
    %\rule{4cm}{3cm}
   \label{fig:copd_nomogram_p3}
   %\caption{Interference frequency $\omega_{\text{int}} \in [0.5,3]$}
}
\caption{The design nomogram for comparator overdrive.}
\end{figure*}
\end{comment}

% \begin{figure}[htp]
%     \subfigure[Overdrive delay --- overshoot tradeoff.]
%     {\centering
%     \includegraphics[width = 7cm]{no_tradeoff_copd_oo.pdf}
%     \label{fig:no_tradeoff_copd_oo}
%     }
%     \centering
%     \subfigure[Overdrive delay --- settling tradeoff.]
%     {\centering
%     \includegraphics[width = 7 cm]{no_tradeoff_copd_settling.pdf}
%     \label{fig:no_tradeoff_copd_settling}
%     }
%     \caption{Transient performance design diagrams for comparator overdrive.\red{put the legend nearby the curves}}
%     \label{fig:c_comp}
% \end{figure}

\begin{figure}[htp]
    \subfigure[LT1711]
    {\centering
    \includegraphics[width = 7cm]{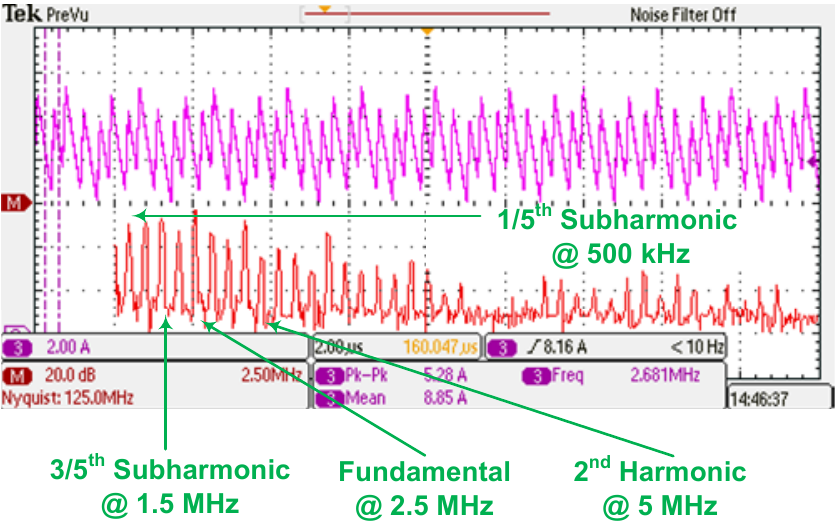}
    \label{fig:complt1711}
    }
    \centering
    \subfigure[AD8469]
    {\centering
    \includegraphics[width = 7 cm]{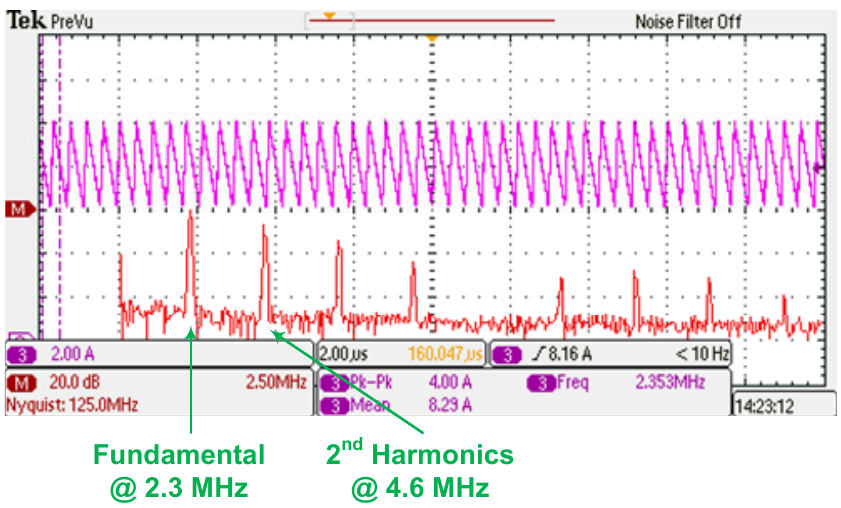}
    \label{fig:comp_da_8469}
    }
    \caption{Fourier transform of the inductor current of the current control loops with different comparators. The current control loop goes unstable when the comparator with shorter overdrive delay (LT1711) is used. The current control loop is stable when the comparator with longer overdrive delay (AD8469) is used.}
    \label{fig:c_comp}
\end{figure}

A neuromorphic analogy to comparator-overdrive-delay conditioning is reminiscent of how integrate-and-fire neurons behave \cite{gerstner2014neuronal}.  This behavior in the control conditioning framework as it pertains to current sensor interference was discussed in Section\,\ref{section:opd}.

The experimental results for the comparator-overdrive-delay conditioning in a buck converter prototype can be observed in Fig.\,\ref{fig:c_comp}. 
%To illustrate the how hardware demonstration of the %comparator,
The current control loop using an LT1711 \cite{Dual2001} whose comparator time constant $\tau_{_\mathcal{C}}$ is too small is unstable.
The current control loop using  AD8469 comparator \cite{DS_AD8469}, whose comparator time constant is high enough, is stable. 

\begin{comment}
The datasheet does not explicitly provide the comparator time constant, but instead it includes the overdrive delay diagram as shown in Figs.\,\ref{fig:simumodelcomp_lt1711} and \ref{fig:simumodelcomp_ad8469}. We fit the data sheet plot by the linear regression equation.
\begin{align}
    t_{od} = p_1 \frac{1}{V_{od}} + p_2.
\end{align}
$p_1$ is the fitted time constant --- trigger voltage product and $p_2$ is the fitted constant propagation delay.
\end{comment}

Figures\,\ref{fig:complt1711} and  \ref{fig:comp_da_8469} show the experimental waveforms of the current\nobreakdash-mode buck converter with different comparator overdrive delays.
The inductor current is shown in channel 3 (pink) and the current sensor output is shown in channel 1 (dark blue), which is highly contaminated by interference.

As discussed in Section\,\ref{sec:Intro}, subharmonics in the switching result in poor converter behavior.
Fig.\,\ref{fig:complt1711} shows the subharmonics when the comparator overdrive delay is too small. 
From the Fourier transform of the inductor current, we observe that in addition to the switching frequency, the $1/5^{\text{th}}$, $2/5^{\text{th}}$, $3/5^{\text{th}}$, $4/5^{\text{th}}$\nobreakdash-order subharmonics are also commingled because of the interference.
Comparator overdrive delay can be increased using a different comparator;
as shown in Fig.\,\ref{fig:comp_da_8469}, the inductor current becomes stable in the periodic steady state. 
When stable, the Fourier transform of the inductor current only contains switching\nobreakdash-frequency and higher\nobreakdash-order harmonics. 

From the  experimental data, the steady state behavior of the current control loop agrees with the theory. A slow comparator can condition the interference and reduce the nonlinearity of current mapping so that the current loop is stable.
\section{Conclusion}\label{sec:conlusion_part2}
In this paper, we presented the analysis and design of three control conditioning methods for extremum current-mode controllers in high-frequency converters with disruptive current sensor interference.  We provided a rigorous model for the dynamics of the current loop using these conditioning methods with interference, to ensure robust stability.  Specifically, we compared and rigorously analyzed: 
(1) comparator overdrive delay conditioning; 
(2) slope compensation; 
and (3) low-pass filter conditioning, within the unified framework in Part 1 of this paper series. 
We experimentally demonstrated and validated in a multi-MHz power converter hardware the effectiveness of all three methods where interference was harsh.

\section*{Acknowledgements}
Special thanks to Professor Peter Seiler for the contribution in large-signal stability analysis. This work is based upon work supported by the U.S. Department of Energy SunShot Initiative, under Award Number(s) DE-EE-0007549.

\newpage
\begin{appendices}

\section{Small-Signal Model of Current Control Loop with Time-Varying Inductor Current Ramp} \label{sec:small_ccl}
If the coupling between the output capacitor voltage and inductor current ripple is not negligible, the small-signal model of current control loop with time-varying inductor current ramp follows: 
\begin{align}
    \frac{\hat{t}_{\text{off}}(z)}{\hat i_v(z)} = g\,\frac{(1-b_1z^{-1}-b_2z^{-2})}{1-a_1z^{-1}},
\end{align}
where
\\
\scalebox{0.9}{\vbox{
\begin{align*}
a_1 &= 1-(1+M_r)\,\hat{\tau}^{-1}_1-\frac{1+M_r}{2}\hat{\tau}^{-1}_1\hat{\tau}^{-1}_2, \nonumber \\
b_1 &= 2 - (1+M_r)\hat{\tau}^{-1} - \left(\frac{(1+M_r)^2}{2} + (1+M_r)\lambda \right)\hat{\tau}^{-1}_1\hat{\tau}^{-1}_2,\nonumber \\
b_2 &= -1 + (1+M_r)\hat{\tau}^{-1} - \left(\frac{(1+M_r)^2}{2} - (1+M_r)\lambda \right)\hat{\tau}^{-1}_1\hat{\tau}^{-1}_2,\nonumber \\
g &= \frac{L}{V_{\text{out}}}, \quad M_r = \frac{m_1}{m_2}, \quad \hat{\tau}_1 = \frac{RC}{T_{\text{on}}}, \quad \hat{\tau}_2 = \frac{L/R}{T_{\text{on}}}.
\end{align*}}}

\section{Considerations for Designing a Voltage Controller Around the Current Control Loop} \label{sec:small_vcl_ccl}
A current-mode voltage converter has a major loop to control the output voltage and a minor loop to control the current. Although current control loop $C(z)$ is the main topic of this paper, the design of the voltage controller depends on the design of the current control loop. To design the voltage controller $K(z)$, we need a model for the converter plant $\Sigma$. The \emph{5S} framework in \cite{Cui2018a} provides a purely digital, simple and accurate model for this type of power converter.
\begin{align}
    \frac{\hat v(z)}{\hat i_v(z)} = g\,\frac{(1-b_1z^{-1})\,z^{-1}}{1-a_1z^{-1}},
\end{align}
where 
\begin{align}
a_1 &= 1-(1+M_r)\,\hat{\tau}^{-1}_1-\frac{1+M_r}{2}\hat{\tau}^{-1}_1\hat{\tau}^{-1}_2, \nonumber \\  g &= R \left(\lambda + \frac{M_r}{2}\right) \hat{\tau}^{-1}_1,\quad
b_1 = -\frac{1-\lambda+ M_r/2}{\lambda+M_r/2},\nonumber \\
M_r & = \frac{m_1}{m_2}, \quad \hat{\tau}_1 = \frac{RC}{T_{\text{on}}}, \quad \hat{\tau}_2 = \frac{L/R}{T_{\text{on}}}.
\end{align}
Given the plant $\Sigma(z)$ and current control loop $C(z)$ in this section, the designer can design the voltage controller $K(z)$ using the root-locus method in \cite{Cui2019c}.
For other power converters like dc-dc converters using fixed-frequency control, the plant model is provided in Appendix \ref{appendix:fpccmodel}.
% The plant model of other power converters like dc-dc converters using fixed-frequency control can be derived similarly.

\section{Sampled-Data Space Modeling of a DC-DC Converter Using Fixed-Frequency Peak Current-Mode Control} \label{appendix:fpccmodel}
We take the boost converter using fixed-frequency current-mode control with parameters 
%in Table \ref{table:fpcc_buck_param} 
as an example. 
The model is derived by following the sampled-data modeling method in \cite{Verghese2003} as:
 \begin{align}
     \frac{\hat v(z)}{\hat i_p(z)} = g \frac{(1 - b_1z^{-1})z^{-1}}{1 - a_1z^{-1}}
 \end{align}
 where
%  \begin{align}
%      g &= -\frac{2 L (M+1)^2+(2 M+1) R T_s}{2 (M+1)^2 RC},\\
%      a_1 &=  -\frac{(2 M+1) T_s \left(2 L (M+1)^2+M R T_s\right)-2 LC (M+1)^3 R}{2 L (M+1)^3 RC}, \\
%      b_1 &= \frac{2 L (M+1)^2+\left(2 M^2+4 M+1\right) R T_s}{2 L (M+1)^2+(2 M+1) R T_s}, \\
%      M & = \frac{m_1}{m_2}.
%  \end{align}
  \begin{align}
     g &= R\left( -\hat{\tau}_2\hat{\tau}^{-1}_1 + \frac{2M_r+1}{2(M_r+1)^2}\hat{\tau}^{-1}_1\right),\\
     a_1 &= 1 - \frac{2M_r+1}{M_r+1}\hat{\tau}^{-1}_1 - \frac{2M_r+1}{(M_r+1)^3}\hat{\tau}_1^{-1}\hat{\tau}_2^{-1}, \\
     b_1 &= \frac{2(M_r+1)^2\hat{\tau}_2 + (2M_r^2+4M_r+1)}{2(M_r+1)^2\hat{\tau}_2 + (2M_r+1)}, \\
     M_r & = \frac{m_1}{m_2}, \quad \hat{\tau}_1 = \frac{RC}{T_s}, \quad \hat{\tau}_2 = \frac{L/R}{T_s}.
 \end{align}

\end{appendices}
\newpage
{
\setstretch{1}\vspace{\baselineskip}
\bibliographystyle{IEEEtran}
\bibliography{main.bib}
}

\end{document}